\documentclass[preprint,showpacs,preprintnumbers,amsmath,floatfix,amssymb]{revtex4}
\newcommand{\newwidth}{0.675\textwidth}
\newcommand{\newheight}{0.45\textwidth}
\newcommand{\newwidthprime}{0.225\textwidth}
\newcommand{\newheightprime}{0.30\textwidth}

\usepackage{graphicx} 
\usepackage{epstopdf}  
\usepackage{float}
\usepackage{color}
\usepackage{dcolumn}
\usepackage{bm,verbatim}
\usepackage{amsmath} 
\usepackage{bm,lineno} 
\usepackage{natbib} 

\begin{document}

\title{Bias-Controlled Selective Excitation of Vibrational Modes in Molecular Junctions: 
A Route Towards Mode-Selective Chemistry}
\author{Roie Volkovich$^{(a)}$}
\author{Rainer H\"artle$^{(b)}$}
\author{Michael Thoss$^{(b)}$}
\author{Uri Peskin$^{(a)}$}
\affiliation{$^{(a)}$Schulich Faculty of Chemistry
and the Lise Meitner Center for Computational Quantum Chemistry,\\
Technion-Israel Institute of Technology, Haifa 32000, Israel\\
$^{(b)}$Institut f\"ur Theoretische Physik und Interdisziplin\"ares Zentrum f\"ur Molekulare Materialien,\\
Friedrich-Alexander-Universit\"at Erlangen-N\"urnberg,
Staudtstr.\,7/B2, D-91058 Erlangen, Germany}

\begin{abstract}
We show that individual vibrational modes in single-molecule junctions with asymmetric 
molecule-lead coupling can be selectively excited by applying an external bias voltage. 
Thereby, a non-statistical distribution of 
vibrational energy can be generated, that is, 
a mode with a high frequency can be stronger excited than a mode with a lower 
frequency. This is of particular interest in the context 
of mode-selective chemistry, where one aims to break specific (not 
necessarily the weakest) chemical bond in a molecule. 
Such Mode-Selective Vibrational Excitation is demonstrated
for two generic model systems representing asymmetric molecular junctions 
and/or Scanning Tunneling Microscopy experiments.   
To this end, we employ two complementary theoretical approaches, 
a nonequilibrium Green's function approach and a 
master equation approach. 
The comparison of both methods reveals good agreement in describing resonant 
electron transport through a single-molecule contact, 
but also highlights the role of non-resonant transport processes, in particular 
co-tunneling and off-resonant electron-hole pair creation processes. 
\end{abstract}

\maketitle

\section{Introduction}

Mode-selective chemistry, that is the control of specific conformational changes
or chemical reactions by directing energy into specific vibrational modes of a
molecule, has been a major goal and challenge of modern chemical physics
\cite{Jortner91,Liu2006,Crim2008}. 
A variety of different routes to achieve mode-selective chemistry have been 
considered. For example, 
ultrafast excitation of molecules by laser pulses provided a route 
for directing energy transiently into specific 
mode-excitation or bond-cleavage in the presence of 
Intramolecular Vibrational Energy Redistribution (IVR) processes \cite{Potter92,Zare98,Brumer03}. 
Nevertheless, on long time scales the statistical distribution of energy usually prevails and the control of specific conformational changes or chemical reactions by directing energy into specific vibrational modes still represents a challenge \cite{Bryan}. 

In recent years, much interest has been devoted to the study of molecular systems out of equilibrium, in particular single-molecule junctions \cite{Joachim97,Park00,Reichert02,Smit02,Nitzan03,Pascual03,WuNazinHo04,Ogawa07,Hihath,Schulze08,Pump08,Ballmann2010,Osorio2010,Tao2010,cuevasscheer2010,Secker2011}. These junctions consist of a single molecule that is clamped between two metal or semi-conductor electrodes. If these electrodes are set to different electrochemical potentials by applying an external bias voltage, electrons tunnel from one lead to the other trough the molecule.  The geometrical  structure of the molecular bridge, due to its small size and mass, is very sensitive to charge fluctuations induced by these tunneling processes. The tunneling electrons thus strongly interact with the vibrational degrees of freedom of the junction \cite{WuNazinHo04,Natelson2004,Huang2006,Natelson2008,Repp,Ballmann2010,Tao2010,Osorio2010,Secker2011}. The distribution of vibrational energy on the molecular bridge, which results from these interactions, is highly correlated with the applied bias voltage and often deviates from a Boltzmann distribution \cite{Mitra04,Semmelhack,Koch05,Natelson2008,Ioffe08,Hartle,Hartle09,MSVE10,Hartle2010b,Wohlman2010,Hartle2011,Hartle2011b}. In a single-molecule junction, it is thus possible to control a "non-statistical" non-equilibrium distribution of vibrational energy by an external potential bias. 
This offers an alternative route to mode-selective excitation 
(or mode-selective chemistry). Moreover, in the steady-state transport 
regime of a molecular junction,  
this could be achieved even without the cumbersome 
preparation of a specific initial-state. 

Employing generic model systems, we have recently shown \cite{MSVE10} that in a single-molecule junction the excitation of vibrational modes can be, indeed, selectively controlled by the external bias voltage.
Considering molecules with an asymmetric orbital structure and vibronic coupling to specific normal modes, it was demonstrated that, by adjusting the external bias voltage, such a system can be driven into different non-equilibrium states 
with different levels of excitation for specific nuclear modes. 
Particularly, the excitation of a high frequency mode can be tuned much higher 
than that of a low frequency mode, thus overcoming the statistical distribution of 
vibrational energy that favors the excitation of low frequency modes. 
In this article, we extend our previous studies \cite{MSVE10}, 
where the principle of Mode-Selective Vibrational Excitation (MSVE) 
in a single-molecule junction was demonstrated for the first time. 
To this end, we outline in detail how the excitation of a single vibrational 
mode can be controlled by an external bias voltage. 
This bias-controlled excitation is then generalized to more than 
one vibrational mode, reviewing the basic MSVE phenomenon 
and emphasizing the role of intra-molecular interactions.  
In particular, we demonstrate 
MSVE in the presence of electronically mediated mode-mode coupling \cite{Hartle}, 
which results from coupling of 
the vibrational modes 
to the same electronic state. 
These interactions induce energy transfer between the vibrational modes and 
tend to distribute current-induced vibrational excitation  
between the different modes.  
Similarly, electronic correlations, 
\emph{e.g.}\ due to Coulomb repulsion, may influence MSVE by reorganizing 
the electronic population 
between specific electronic states at the molecule, which,   
to some extent and in different ways, is related to the phenomenon of MSVE. 
Extending our previous work, we investigate complementary 
models for asymmetric molecular junctions exhibiting MSVE. 
These models are related, for example, to experiments 
on single-molecule junctions performed with a Scanning Tunneling Microscope (STM) \cite{Gaudioso00,Pascual03,Choi,Ogawa07,Pump08,Lafferentz2009,Schulze08,Repp}.

To describe this nonequilibrium transport problem, 
we employ two complementary approaches. The first one is a Master Equation approach (ME) that is based on 
the time-evolution of the reduced density matrix of the molecular bridge \cite{May02,Mitra04,Lehmann04,Neuhauser05,Pedersen05,Welack06,Harbola2006,Zazunov06,Siddiqui,
Timm08,May08,May08b,Leijnse09,Esposito09,Esposito2010,Peskin2010,MSVE10,Hartle2010b,Cizek2011,Volkovich2011,Hartle2011}. 
The respective equation of motion is evaluated strictly to second-order in the molecule-lead coupling, for which 
all resonant transport processes are included. 
Within such a framework, no approximations with respect to the interactions 
on the molecular bridge need to be invoked, in particular not with respect to
electron-electron interactions or electronic-vibrational coupling. 
However, higher-order processes \cite{Koenig96,Paaske2005,Sela,Galperin2007,Lueffe,Hartle,Leijnse2009,Han2010} 
like co-tunneling processes or the broadening 
of molecular levels due to the molecule-lead coupling are missing in this description. 
Note that master equation approaches 
that take into account such higher-order effects have already 
been put forward \cite{May02,Pedersen05,Leijnse09,Esposito09,Esposito2010}.  
In the present work, however, we selected a Nonequilibrium Green's Function approach (NEGF)
\cite{Flensberg03,Mitra04,Galperin06,Ryndyk06,Frederikesen07b,Tahir2008,Hartle,Avriller2009,Schmidt2009,Haupt2009,Stafford09,Hartle09,Kubala2010,MSVE10,Hartle2011b} 
to account for the higher-order effects. 
Especially for the description of multiple vibrational modes \cite{Hartle,MSVE10,Secker2011}, 
the NEGF methodology is typically more efficient.  
We employ a nonequilibrium Green's 
function approach, which was originally proposed by 
Galperin \emph{et al.} \cite{Galperin06} and 
recently extended to account for multiple vibrational and 
multiple electronic degrees of freedom of the molecular bridge \cite{Hartle,Hartle09,MSVE10,Hartle2011b}. 
Other theoretical approaches used to describe electron transport through a single-molecule junction  
are based \emph{e.g.}\ on scattering theory 
\cite{Cizek04,Cizek05,Seidemann05,Toroker07,Benesch08,Zimbovskaya09,Seidemann10}, path integrals \cite{Muehlbacher08,Thorwart2008,Segal2010}, 
multiconfigurational wave-function methods \cite{Wang09,Velizhanin2010,Thoss2011}, flux-correlation approaches \cite{Maytal10} 
or exact diagonalization \cite{Hackl10}. 

The paper is organized as follows. The Hamiltonian that we use to describe electron transport through 
a single-molecule junction is outlined in Sec.\ \ref{modHam}. The ME and NEGF methodologies that we employ to 
calculate steady-state observables of a biased single-molecule junction are briefly described in 
Secs.\ \ref{MEapproach} and \ref{NEGFapproach}, respectively. The basic physical mechanisms for 
vibrational heating and cooling in an asymmetric molecular junction, 
which lead to MSVE, are reviewed and 
analyzed for a single-level conductor in Sec.\ \ref{VSE}. Thereby, the role of electron-hole 
pair creation processes, which constitutes an important cooling mechanism 
in a molecular junction \cite{MSVE10,Hartle2010b,Hartle2011}, is discussed in detail. 
Moreover, the 
comparison of results obtained from our NEGF and ME schemes enables us to distinguish 
resonant and non-resonant contributions. 
In Sec.\ \ref{MSVE}, we discuss MSVE in two model systems, representing  
generic asymmetric molecular junctions: A single-molecule junction with an 
intrinsically asymmetric molecular bridge (model A) \cite{MSVE10},  
and a junction, where the bridging molecule is asymmetrically coupled to the leads (model B). 
The latter scenario is typical for STM experiments. 
We demonstrate that 
the magnitude and the polarity of the external bias voltage can be used to direct vibrational energy 
into specific vibrational modes, 
even in the presence of intra-molecular interactions that tend to suppress the effect. 
Thus, we predict that in the steady-state transport regime of a molecular junction, 
a bias-controlled "non-statistical" distribution of vibrational energy can be realized, 
where modes with higher frequency (stronger bonds) can be much higher 
excited than modes with a lower frequency (weaker bonds).

\section{Theoretical Methodology}

\subsection{Model Hamiltonian}
\label{modHam}

We consider electron transport through a single molecule that is covalently bound to two metal leads. 
To this end, we employ a generic model Hamiltonian 
\begin{eqnarray}
\label{H}
\hat{H} &=& \hat{H}_{\text{el}} + \hat{H}_{\text{vib}},   
\end{eqnarray}
describing the electronic, $\hat{H}_{\text{el}}$, and the vibrational degrees of freedom, $\hat{H}_{\text{vib}}$, of this transport problem. 

The electronic part of $\hat{H}$ can be represented by a discrete set of electronic states, located at the molecular bridge (M),  
and by a continuum of electronic states describing the electron reservoirs of the left (L) and the right (R) electrode. 
Tunneling of electrons from one lead to the other is described by the following model Hamiltonian ($\hbar=1$):  
\begin{eqnarray}\label{eq.1}
  \hat{H}_{\text{el}}&=&\sum_{m\in \text{M}}{\epsilon_m a_m^{\dagger}a_m} +\sum_{m<n\in \text{M}}{U_{m,n} a_m^{\dagger}a_m a_n^{\dagger}a_n}  \\
  &&+\sum_{k\in \text{L,R}}{\epsilon_k b_k^{\dagger}b_k} +\sum_{K \in \text{L,R};k \in K;m \in \text{M}}{(\upsilon_{K,m}\xi_{K,k} b_k^{\dagger}a_m+h.c.)}. \nonumber
\end{eqnarray}
The energies of the electronic states in the leads, which are addressed by creation and annihilation operators $b_{k}^{\dagger}$ and $b_{k}$, are denoted by $\epsilon_{k}$. The energy of the $m$th electronic state  located at the molecular bridge is given by $\epsilon_{m}$. These states are populated with creation operators $a_{m}^{\dagger}$ and depopulated with annihilation operators $a_{m}$. The molecular bridge is bilinearly coupled to the electrodes, $\sim b_{k}^{\dagger}a_{m}$, with coupling strengths $\upsilon_{K,m}\xi_{K,k}$. For simplicity, these coupling strengths are factorized into an electrode term,  $\xi_{K,k}$, which determines the electrode's ($K$=L,R) spectral density, $J_{K}(\epsilon)=\sum_{k\in K} \vert\xi_{K,k}\vert^{2} \delta(\epsilon-\epsilon_{k})$, and a molecular term, which represents the coupling of the $m$th molecular state to the $K$th electrode, $\upsilon_{K,n}$. To model the leads we use a semi-elliptic conduction-band with a band-width of $4\gamma$ such that the corresponding level-width functions read: 
\begin{eqnarray}
 \Gamma_{K,mn}(\epsilon)=2\pi \upsilon_{K,m}\upsilon_{K,n}J_{K}(\epsilon)=\upsilon_{K,m}\upsilon_{K,n}\frac{\xi^2}{{\gamma}^2}\sqrt{4\gamma^2-{(\epsilon-\mu_K)}^2}. 
\end{eqnarray}
Charging energies, \emph{e.g.}\ due to Coulomb interactions, are accounted for by Hubbard-like electron-electron interaction terms, $U_{m,n}a_{m}^{\dagger}a_{m} a_{n}^{\dagger}a_{n}$. 
The Fermi-energy of the overall system is given by $\epsilon_{\text{F}}=0$\, eV.

Vibrational degrees of freedom of the molecule are described as harmonic oscillators, 
\begin{eqnarray}
\label{eq.2}
\hat{H}_{\text{vib}} &=& \sum_{\nu} \Omega_{\nu} c_{\nu}^{\dagger}c_{\nu}
+ \sum_{m\in\text{M};\nu} \lambda_{\nu, m} q_{\nu} a_{m}^{\dagger}a_{m} \nonumber\\
&& + \sum_{\nu,\beta_\nu} \omega_{\beta_\nu} d_{\beta_\nu}^{\dagger}d_{\beta_\nu} + \sum_{\nu,\beta_\nu}  \eta_{\nu,\beta_{\nu}} (d_{\beta_\nu}^{\dagger}+d_{\beta_\nu}) (c_{\nu}^{\dagger}+c_{\nu}) ,
\end{eqnarray}
where the ladder operators $c^{\dagger}_{\nu}$/$c_{\nu}$ 
address the $\nu$th vibrational (normal) mode of the molecular bridge with  
frequency $\Omega_{\nu}$. 
Changes in the nuclear potential energy surface due to electronic transitions between 
the single particle states are assumed to be linear in both the vibrational 
displacements, $q_{\nu}=\frac{1}{\sqrt{2}}(c_{\nu}+c_{\nu}^{\dagger})$, and the 
electronic densities, $a_{m}^{\dagger}a_{m}$. The respective coupling strengths are given 
by $\lambda_{\nu, m}$. 
To incorporate vibrational relaxation effects in a phenomenological way \cite{Weiss93,Thoss98}, 
each intramolecular vibrational mode is coupled to a thermal bath. 
The creation and annihilation operator for a bath mode with frequency 
$\omega_{\beta_{\nu}}$ are denoted by $d^\dagger_{\beta_{\nu}}$ and $d_{\beta_{\nu}}$, respectively. 
The corresponding mode-bath coupling strengths are given by $\eta_{\nu,\beta_{\nu}}$. All properties of the bath 
that influence the dynamics of the system are determined by the spectral densities 
$J_{\nu}(\omega)= \sum_{\beta_{\nu}}\eta_{\nu,\beta_{\nu}}^{2}\delta(\omega- \omega_{\beta_{\nu}})$. 
In the calculations presented below, we use an Ohmic bath model with a cutoff frequency, $\omega_{c,{\nu}}$, and 
\begin{eqnarray}
 J_{\nu}(\omega)= \frac{\zeta_\nu^2} {\omega_{c,\nu}^2}\omega e^{-\omega/\omega_{c,\nu}}. 
\end{eqnarray}

\subsection{Reduced Density Matrix Approach}
\label{MEapproach}

Weak coupling of the molecule to the electron reservoirs (the leads), 
as well as to the energy reservoirs (the nuclear bath modes), 
is essential in order to relate the junction's steady-state observables 
to intrinsic properties of the molecular bridge \cite{Maytal2009}. 
For this purpose it is instructive to regroup the different terms in 
the Hamiltonian into the molecular "system", which include the molecular 
electronic states and vibrational modes, and "bath" terms, which include 
the electron reservoirs and the nuclear baths. 
The full Hamiltonian is therefore rewritten as: 
\begin{eqnarray}\label{eq.3}
\hat{H} &=& \hat{H}_{\text{S}} + \hat{H}_{\text{B}}+ \hat{H}_{\text{SB}},
\end{eqnarray}
with 
\begin{eqnarray}\label{eq.4}
  \hat{H}_{\text{S}}=&&\sum_{m\in \text{M}}{\epsilon_m a_m^{\dagger}a_m}+\sum_{m<n\in \text{M}}{U_{m,n} a_m^{\dagger}a_m a_n^{\dagger}a_n} + \sum_{\nu}{\Omega_\nu c_{\nu}^{\dagger}c_{\nu}} +\sum_{m\in \text{M};\nu}{\lambda_{\nu, m} q_{\nu} a_m^{\dagger}a_m},  \\
  \hat{H}_{\text{B}}=&&\sum_{k\in \text{L,R}}{\epsilon_k b_k^{\dagger} b_k}+\sum_{\nu,\beta_{\nu}}{\omega_{\beta_{\nu}} d_{\beta_{\nu}}^{\dagger}d_{\beta_{\nu}}}, \\
  \hat{H}_{\text{SB}}=&&\sum_{\nu,\beta_{\nu}}{\eta_{\nu,\beta_{\nu}}(d_{\beta_{\nu}}^{\dagger}+d_{\beta_{\nu}})(c_{\nu}^{\dagger}+c_{\nu})} +\sum_{K \in \text{L,R};k \in K;m \in \text{M}}{(\upsilon_{K,m}\xi_{K,k}. b_k^{\dagger}a_m+h.c.)} 
\end{eqnarray}
The steady-state of the system under bias can be calculated by following the time-evolution of the system 
to its stationary state, starting from an arbitrary initial state. 
We consider an initial density matrix in a product form, 
\begin{eqnarray}\label{eq.5}
  \hat{\rho}(0)&=&\hat{\rho}_{\text{S}}(0)\otimes\hat{\rho}_{\text{B}}(0), \\
 \hat{\rho}_{\text{B}}(0) &=& \prod_{K \in \text{L,R}}{\hat{\rho}_{\text{B},K}}\otimes\prod_{\nu}{\hat{\rho}_{\text{B},\nu}}, 
\end{eqnarray}
where $\hat{\rho}_{\text{S}}(0)$ is any normalized system density matrix with $\text{tr}_{\text{S}}[\hat{\rho}_{\text{S}}(0)]=1$, and 
$\text{tr}_{\text{S}}[...]$ denotes the trace over the subspace of the molecular conductor. 
The electronic reservoirs are described by 
a product of equilibrium density operators, 
\begin{eqnarray}\label{eq.6}
  \hat{\rho}_{\text{B},K}=\frac{e^{-\frac{1}{k_{\text{B}}T}\sum_{k\in K}{(\epsilon_k-\mu_K)b^{\dagger}_k b_k}}}{\text{tr}[e^{-\frac{1}{k_{\text{B}}T}\sum_{k\in K}{(\epsilon_k-\mu_K)b^{\dagger}_k b_k}}]},
\end{eqnarray}
with $K\in{\text{L,R}}$. Thus, we assume the leads to be in a thermal equilibrium state, which is 
characterized by the temperature $k_{\text{B}}T$ and the electrochemical potentials $\mu_K$. 
Similarly, the nuclear baths are represented by 
a product of equilibrium density operators, 
\begin{eqnarray}\label{eq.7}
  \hat{\rho}_{\text{B},\nu}=\frac{e^{-\frac{1}{k_{\text{B}}T}\sum_{\beta_{\nu}}{\omega_{\beta_{\nu}}d^{\dagger}_{\beta_{\nu}} d_{\beta_{\nu}}}}}{\text{tr}[e^{-\frac{1}{k_{\text{B}}T}\sum_{\beta_{\nu}}{\omega_{\beta_{\nu}}d^{\dagger}_{\beta_{\nu}} d_{\beta_{\nu}}}}]}, 
\end{eqnarray}
where $\nu$ denotes the molecular mode to which the particular bath is coupled. 
The exact time-evolution of the full density operator is given by the 
Liouville-von Neumann equation, 
\begin{eqnarray}\label{eq.8}
  \frac{\partial}{\partial t}\hat{\rho}(t)=-i[\hat{H},\hat{\rho}(t)].
\end{eqnarray}
Transforming the respective operators to the interaction representation, 
\begin{eqnarray}
 \hat{O}^\text{I}(t)=e^{i[\hat{H}_{\text{S}}+\hat{H}_{\text{B}}]t}\hat{O}(t)e^{-i[\hat{H}_{\text{S}}+\hat{H}_{\text{B}}]t},
\end{eqnarray}
the Liouville-von Neumann equation can be rearranged to give \cite{Peskin2010}
\begin{eqnarray}\label{eq.9}
  \frac{\partial}{\partial t}\hat{\rho}^\text{I}(t)=-i[\hat{H}^\text{I}_{\text{SB}}(t),
  \hat{\rho}^\text{I}(t)]-\int_0^{t}{\text{d}t'[\hat{H}^\text{I}_{\text{SB}}(t),[\hat{H}^\text{I}_{\text{SB}}(t'),
  \hat{\rho}^\text{I}(t)]]} \nonumber\\
  -i\int_0^{t}{\text{d}t'\int_{t'}^{t}{\text{d}t''[\hat{H}^\text{I}_{\text{SB}}(t),[\hat{H}^\text{I}_{\text{SB}}(t'),[\hat{H}^\text{I}_{\text{SB}}(t''),
  \hat{\rho}^\text{I}(t'')]]]}}.
\end{eqnarray}
Assuming weak system-bath coupling, the third term in Eq.\ (\ref{eq.9}) can be 
neglected, 
which yields  
\begin{eqnarray}\label{eq.10}
  \frac{\partial}{\partial t}\hat{\rho}^\text{I}(t)&\cong&-i[\hat{H}^\text{I}_{\text{SB}}(t),
  \hat{\rho}^\text{I}(t)] -\int_0^{\infty}{\text{d}t'[\hat{H}^\text{I}_{\text{SB}}(t),[\hat{H}^\text{I}_{\text{SB}}(t'),
  \hat{\rho}^\text{I}(t)]]}. 
\end{eqnarray}
Defining the reduced density operator $\hat{\rho}^\text{I}_{\text{S}}(t)\equiv \text{tr}_{\text{B}}[\hat{\rho}^\text{I}(t)]$, 
one obtains the well-established (Markovian) Master equation for $\hat{\rho_{\text{S}}}(t)$  \cite{Redfield1965,Blum,Egorova03,May04,Harbola2006,Volkovich2008,Hartle2010b,Volkovich2011,Hartle2011} 
by replacing 
$\hat{\rho}^\text{I}(t)$ by $\hat{\rho}_{\text{B}}(0)\otimes \hat{\rho}^\text{I}_{\text{S}}(t)$ 
in Eq.\ (\ref{eq.10}), and taking the integration limit to infinity, $\int_{0}^{t} \to \int_{0}^{\infty}$,
\begin{eqnarray}\label{eq.11}
  &&\frac{\partial}{\partial t}\hat{\rho}_{\text{S}}(t)=-i[\hat{H}_{\text{S}},
  \hat{\rho}_{\text{S}}(t)] - \int_0^{\infty}{\text{d}t' \text{tr}_{\text{B}}[\hat{H}_{\text{SB}},[\hat{H}_{\text{SB}}(t-t'),
  \hat{\rho}_{\text{B}}(0)\hat{\rho}_{\text{S}}(t)]]},
\end{eqnarray}
with  $\hat{H}_{\text{SB}}(\tau)=e^{-i(\hat{H}_{\text{S}}+\hat{H}_{\text{B}})\tau}\hat{H}_{\text{SB}}e^{i(\hat{H}_{\text{S}}+\hat{H}_{\text{B}})\tau}$. Thereby, we used that 
$[ \hat{\rho}_{\text{B}}(0), \hat{H}_{\text{B}} ] =0$ 
and $\text{tr}_{\text{B}}[ \hat{\rho}_{\text{B}}(0) \hat{H}_{\text{SB}} ] =0$ \cite{Peskin2010}. 
To evaluate this equation of motion, it is convenient to use 
the eigenstates of the molecular system Hamiltonian, 
\begin{eqnarray}\label{eq.12}
 \hat{H}_{\text{S}}|l\rangle=E_l|l\rangle.  
\end{eqnarray}
Taken in this basis, Eq.\ (\ref{eq.11}) corresponds 
to the Redfield equation \cite{Redfield1965,Blum,May04}. 
The molecular system observables at steady-state can be calculated from the 
infinite time limit of $\hat\rho_{\text{S}}(t)$. 
Moreover, the effects of coherences between the system eigenstates can be neglected in this limit 
as long as the 
molecular levels are non-degenerate \cite{Hartle2010b} (Coherences between quasi-degenerate molecular levels 
can play an important role, as in the case for molecular motors \cite{Cizek2011}).  
For the model systems that we study in Sec.\ \ref{results} coherences are not important 
in the steady state limit and  
consequently, the equation of motion for the diagonal matrix elements of the reduced density matrix, 
that is the populations $P_{l}(t)=\hat{\rho}_{\text{S},l,l}(t)$, is given by 
\begin{eqnarray}\label{eq.13}
\frac{\partial}{\partial t} P_{l}(t)=\sum_{l'}[\kappa^{\text{R}}+\kappa^{\text{L}}+\sum_{\nu}{\kappa^{(\nu)}}]_{l,l'} P_{l'}(t).    
\end{eqnarray}
The respective rate matrices for electron tunneling take the form, 
\begin{eqnarray}\label{eq.14}
[\kappa^{\text{L/R}}]_{l,l'}&=&(1-\delta_{l,l'})(\Gamma_{l,l'}^{\text{L/R};h}+\Gamma_{l',l}^{\text{L/R};e}) -\delta_{l,l'}\sum_{l'\neq l}{(\Gamma_{l',l}^{\text{L/R};h}+\Gamma_{l,l'}^{\text{L/R};e})},
\end{eqnarray}
and are determined by the spectral densities 
and the Fermi occupation numbers at each electrode,
\begin{eqnarray}\label{eq.15}
\Gamma_{l,l'}^{\text{L/R};\text{e/h}}\equiv 2\pi |\sum_m{\upsilon_{\text{L/R},m}[a_m^{\dagger}]_{l',l}}|^2 J_{\text{L/R}}(E_{l'}-E_{l}) f_{\text{e/h}}^{\text{L/R}}(E_{l'}-E_{l}),
\end{eqnarray}
with
\begin{eqnarray}\label{eq.16}
f_{\text{e}}^{\text{L/R}}(\epsilon)&=&\frac{1}{1+e^{(\epsilon-\mu_\text{L/R})/k_{\text{B}}T}},\\ 
f_{\text{h}}^{\text{L/R}}(\epsilon)&=&1-f_{\text{e}}^{\text{L/R}}(\epsilon). 
\end{eqnarray}
Similarly, the rate matrices describing the coupling of the vibrational modes 
to their thermal bath take the form,
\begin{eqnarray}\label{eq.17}
[\kappa^{\nu}]_{l,l'}&=&(1-\delta_{l,l'})(\Gamma_{l,l'}^{\nu;\text{d}}+\Gamma_{l',l}^{\nu;\text{u}}) - \delta_{l,l'}\sum_{l'\neq l}{(\Gamma_{l',l}^{\nu;\text{d}}+\Gamma_{l,l'}^{\nu;\text{u}})},
\end{eqnarray}
and are accordingly determined by the spectral densities and 
the phonon occupation numbers for each system mode, 
\begin{eqnarray}\label{eq.18}
\Gamma_{l,l'}^{\nu;\text{u/d}}=2\pi|[c_{\nu}^{\dagger}+c_{\nu}]_{l',l}|^2 J_{(\nu)}(E_{l'}-E_{l}) n_{\text{u/d}}(E_{l'}-E_{l}), 
\end{eqnarray}
with
\begin{eqnarray}
n_{\text{u}}(\epsilon)=\frac{1}{e^{\epsilon/k_{\text{B}}T}-1}, \\ n_{\text{d}}=(\epsilon)=n_{\text{u}}(\epsilon)+1. 
\end{eqnarray}

Observables of interest, such as the steady-state current from left to right,
\begin{eqnarray}\label{eq.19}
I_{\text{L}\rightarrow \text{R}}=\lim_{t\rightarrow \infty}\sum_l{\sum_{l'}{2e[\kappa_{l,l'}^{\text{L}}] P_{l'}(t) n_l}},
\end{eqnarray}
the average level of excitation of mode $\nu$,
\begin{eqnarray}\label{eq.20}
\langle c_{\nu}^{\dagger}c_{\nu}\rangle=\lim_{t\rightarrow \infty}\sum_l{P_{l}(t) \langle l |c_{\nu}^{\dagger}c_{\nu}| l \rangle},
\end{eqnarray}
and the populations of the electronic states, 
\begin{eqnarray}\label{eq.21}
\langle a_{m}^{\dagger}a_{m}\rangle=\lim_{t\rightarrow \infty}\sum_l{P_{l}(t) \langle l |a_{m}^{\dagger}a_{m}| l \rangle},
\end{eqnarray}
are calculated from the infinite time limit of the $P_{l}(t)$. 
Thereby, $n_l$ is given by $n_l=\sum_{m} \langle l |a_{m}^{\dagger}a_{m}| l \rangle $. 

\subsection{Nonequilibrium Green's Function Approach}
\label{NEGFapproach}

Alternatively, vibrationally coupled electron transport 
through a single-molecule junction can be described 
employing a nonequilibrium Green's function approach. Using such an approach 
facilitates the description of  
higher-order effects by the associated Dyson-Keldysh equations. 
The comparison of results obtained from the reduced density matrix approach 
and the nonequilibrium Green's function approach  
allows to elucidate the role of these effects 
in vibrationally coupled electron transport through a single-molecule junction. 
Here, we apply the method originally proposed by Galperin \emph{et al.}\ \cite{Galperin06}, which 
we have recently extended to account for multiple vibrational modes and 
multiple electronic states \cite{Hartle,Hartle09,MSVE10,Hartle2011b}. 
The approach is based on the small polaron transformation of the Hamiltonian $\hat{H}$ \cite{Mahan81,Mitra04,Hartle}
\begin{eqnarray}
\label{transformedHamiltonian}
\bar{H} &=& \text{e}^{S} \hat{H} \text{e}^{-S}  \\
&=& \sum_{m} \bar{\epsilon}_{m} a_{m}^{\dagger} a_{m}
+ \sum_{\nu} \Omega_{\nu} c^{\dagger}_{\nu}c_{\nu} + \sum_{n<m} \bar{U}_{m,n} a_{m}^{\dagger}a_{m} a^{\dagger}_{n}a_{n} \\
&&+ \sum_{k} \epsilon_{k} b_{k}^{\dagger}b_{k} + \sum_{\nu,\beta_\nu} \omega_{\beta_\nu} d_{\beta_\nu}^{\dagger}d_{\beta_\nu} \nonumber\\
&&+ \sum_{k\in\text{L,R};m\in\text{M}} ( \upsilon_{K,m}\xi_{K,k} X_{m}
b_{k}^{\dagger}a_{m} + \text{h.c.} ) +\sum_{\nu,\beta_\nu} \eta_{\nu,\beta_{\nu}} (d_{\beta_\nu}^{\dagger}+d_{\beta_\nu}) (c_{\nu}^{\dagger}+c_{\nu}), \nonumber
\end{eqnarray}
with
\begin{eqnarray}
S &=& -i \sum_{m\nu} \frac{\lambda_{\nu, m}}{\Omega_{\nu}} a^{\dagger}_{m}a_{m}
p_{\nu}, \\
X_{m} &=& \text{exp}[i\sum_{\nu}\frac{\lambda_{\nu, m}}{\Omega_{\nu}}
p_{\nu}],\\
p_{\nu} &=& \frac{-i}{\sqrt{2}}\left(c_{\nu}-c_{\nu}^{\dagger}\right).
\end{eqnarray}
The transformed Hamiltonian, $\bar{H}$, thus contains no direct electronic-vibrational coupling term, but  
polaron shifted state-energies ${\bar{\epsilon}_{m}=\epsilon_{m}-\sum_{\nu}
(\lambda_{\nu, m}^{2}/\Omega_{\nu})}$, 
vibrationally induced electron-electron interactions, 
${\bar{U}_{m,n}=U_{m,n}-2\sum_{\nu}(\lambda_{\nu, m}
\lambda_{\nu, n}/\Omega_{\nu})}$,
and shift operators $X_{m}$ that renormalize the molecule-lead 
coupling term. 
Note that 
in Eq.\ (\ref{transformedHamiltonian}) we have neglected the renormalization of the 
molecule-lead coupling term due to coupling of the 
vibrational modes to the thermal baths \cite{Galperin06}. 
Furthermore, the renormalization of the electron-electron interaction terms, 
$\bar{U}_{m,n}a^{\dagger}_{m}a_{m}a^{\dagger}_{n}a_{n}$, 
due to these interactions are also discarded. 
Such bath-induced renormalizations are 
beyond the scope of this paper. 
Also note that the small polaron transformation 
does not allow for an arbitrarily strong coupling 
between the vibrational and the bath modes \cite{Galperin06}, 
which means for the given spectral densities that 
$\zeta_{\nu}^{2}<\Omega_{\nu} \omega_{c,\nu}/4$. 

The single-particle 
Green's function $G_{m,m'}(\tau,\tau')$ is the central quantity 
of (nonequilibrium) Green's function theory. 
With this Green's function all single-particle observables, 
\emph{e.g.}\ the population of levels 
or the current through a single-molecule junction,  
can be readily calculated. 
For the computation of the single-particle Green's function 
$G_{m,m'}(\tau,\tau')$ we employ the following ansatz 
\cite{Galperin06,Hartle,Hartle09,MSVE10,Hartle2011b}:
\begin{eqnarray}
\label{decoupling}
G_{m,m'}(\tau,\tau') &=& -i \langle \text{T}_{c}
a_m(\tau)a_{m'}^\dagger(\tau') \rangle_{\hat{H}} \\
&=&-i \langle \text{T}_{c} a_m(\tau)X_m(\tau)a_{m'}^\dagger(\tau') X^\dagger_{m'}(\tau')\rangle_{\bar{H}} \\
&\approx& \bar{G}_{m,m'}(\tau,\tau') \langle \text{T}_{c} X_m(\tau)X_{m'}^\dagger(\tau') \rangle_{\bar{H}}, \qquad
\end{eqnarray}
with the electronic Green's function $\bar{G}_{m,m'}(\tau,\tau')= -i \langle \text{T}_{c} a_m(\tau)a_{m'}^\dagger(\tau') \rangle_{\bar{H}}$ and  $\text{T}_{c}$ the time-ordering operator on the Keldysh contour.  
The indices $\hat{H}/\bar{H}$ indicate the Hamiltonian, 
which is used to evaluate the respective expectation values. 
The factorization of the Green's function $G_{m,m'}$ into 
a product of an electronic correlation function, $\bar{G}_{m,m'}$, 
and a correlation function of shift operators, $\langle \text{T}_{c} X_m(\tau)X_{m'}^\dagger(\tau') \rangle_{\bar{H}}$, 
is justified, if the dynamics of the electronic and 
the vibrational degrees of freedom are decoupled. 
This is conceptually similar to the 
Born-Oppenheimer approximation \cite{Born1927,Domcke04}. 
Accordingly, for transport through a single-molecule junction, 
an (anti-)adiabatic regime is defined by $\Gamma_{K,mm}\gg\Omega$ 
($\Gamma_{K,mm}\ll\Omega$). 

The self-energy matrices for the electronic part of the Green's function 
can be determined from the equation of motion 
\begin{eqnarray}
 (i\partial_{\tau}-\bar{\epsilon}_{m}) \bar{G}_{m,m'}(\tau,\tau') (-i\partial_{\tau'}-\bar{\epsilon}_{m}) &=& \delta(\tau,\tau') (-i\partial_{\tau'}-\bar{\epsilon}_{m}) \\ 
&&+ \Sigma_{\text{L},m,m'}(\tau,\tau')+ \Sigma_{\text{R},m,m'}(\tau,\tau')+ \Sigma_{\text{Coul},m,m'}(\tau,\tau'). \nonumber
\end{eqnarray}
Here, self-energy contributions due to the coupling 
of the molecule to the left and the right leads are denoted by 
$\Sigma_{\text{L},m,m'}(\tau,\tau')$ and $\Sigma_{\text{R},m,m'}(\tau,\tau')$, while correlations
that result from the electron-electron interaction term, 
$\bar{U}_{m,n}a^{\dagger}_{m}a_{m}a^{\dagger}_{n}a_{n}$, are summarized in 
$\Sigma_{\text{Coul},m,m'}(\tau,\tau')$. 
We treat this latter part of the self-energy in terms of the elastic 
co-tunneling approximation \cite{Groshev,Stafford09,Hartle09,Hartle2011b}. 
We therefore approximate 
$\Sigma_{\text{Coul},m,m'}(\tau,\tau')$ by the self-energy
$\Sigma^{0}_{\text{Coul},m,m'}(\tau,\tau')$, which describes 
electron-electron interactions in the isolated molecule exactly. 
The self-energies that describe the coupling of the molecular bridge to the leads, 
\begin{eqnarray}
\Sigma_{\text{L/R},m,m'}(\tau,\tau')=\sum_{k\in \text{L/R}}\upsilon_{\text{L/R},m}\upsilon_{\text{L/R},m'}|\xi_{\text{L/R},k}|^2 g_k(\tau,\tau')\langle \text{T}_{c} X_{m'}(\tau')X_{m}^\dagger(\tau) \rangle_{\bar{H}},
\end{eqnarray}
are evaluated up to second order 
in the molecule-lead coupling, where $g_k(\tau,\tau')$ denotes the free Green's function 
associated with lead state $k$. 
The real-time projections of these self-energies 
determine the electronic part of 
the single-particle Green's function. 
In the energy-domain the corresponding Dyson-Keldysh equations read 
\begin{eqnarray}
\label{elDy}
 G^{\text{r/a}}_{m,m'}(\epsilon) &=& G^{0,\text{r/a}}_{m,m'}(\epsilon) + \sum_{n,n'} G^{0,\text{r/a}}_{m,n}(\epsilon) \left(\Sigma^{\text{r/a}}_{\text{L},n,n'}(\epsilon) + \Sigma^{\text{r/a}}_{\text{R},n,n'}(\epsilon)\right) G^{\text{r/a}}_{n',m'}(\epsilon),\\
\label{elKe}
 G^{</>}_{m,m'}(\epsilon) &=& \sum_{n,n'} G^{\text{r}}_{m,n}(\epsilon) \left(\Sigma^{</>}_{\text{L},n,n'}(\epsilon) + \Sigma^{</>}_{\text{R},n,n'}(\epsilon)\right) G^{\text{a}}_{n',m'}(\epsilon),\\
\end{eqnarray}
with
\begin{eqnarray}
 G^{0,\text{r/a}}_{m,m'}(\epsilon) &=& g^{0,\text{r/a}}_{m,m'}(\epsilon) + \sum_{n,n'} g^{0,\text{r/a}}_{m,n}(\epsilon) \Sigma^{0,\text{r/a}}_{\text{Coul},n,n'}(\epsilon) G^{0,\text{r/a}}_{n',m'}(\epsilon), \\
g^{0,\text{r/a}}_{m,m'}(\epsilon) &=&  \delta_{m,m'} \frac{1}{\epsilon-\epsilon_{m}+i 0^{+}}.
\end{eqnarray}
For the computation of 
\begin{eqnarray}
 G^{0,\text{r/a}}_{m,m'}(\epsilon) &=& \delta_{m,m'} \sum_{\alpha=1..2^{\text{dim}(\text{M})}} \left(\prod_{m\in\text{M}} (1-n_{m})^{1-(\textbf{p}_{\alpha})_{m}} \ n_{m}^{(\textbf{p}_{\alpha})_{m}} \right) \frac{1}{\epsilon-\epsilon_{m} - \sum_{n} U_{mn}(\textbf{p}_{\alpha})_{n}} 
\end{eqnarray}
we use the populations of the electronic levels
\begin{eqnarray}
 n_{m} &=& \text{Im}\left[ \bar{G}^{<}_{m,m}(\tau=0) \right],   
\end{eqnarray}
that we determine self-consistently, 
and vectors $\textbf{p}_{\alpha}$ that 
point to the edges of a $\text{dim}(\text{M})$-dimensional unit cube. 

Eqs.\ (\ref{elDy}) and (\ref{elKe}) give the exact 
result in the non-interacting limit, 
where $\lambda_{\nu,m}\rightarrow0$ and 
$\bar{U}_{m,n}\rightarrow0$. 
One should bear in mind, however, that 
the elastic co-tunneling approximation treats 
the eigenstates of $\bar{H}$ 
effectively as independent transport channels. 
This description therefore needs to be applied with care, 
if these channels cannot be treated independently from each other, 
\emph{e.g.}\ in the presence of quantum interference effects  \cite{Solomon2006,Hod2006,Solomon2008,Brisker2008,Darau2009,Markussen2010,Hartle2011b}. 
Moreover, Kondo physics \cite{Koenig96,Paaske2005,Sela,Galperin2007,Han2010} 
is also not included in this description, as it employs 
a (self-consistent) second-order expansion in the molecule-lead couplings. 

The correlation function of the shift operators 
is obtained using a second-order cumulant expansion in the dimensionless coupling parameters $\frac{\lambda_{\nu, m}}{\sqrt2\Omega_\nu}$ \cite{Galperin06,Hartle,MSVE10}
\begin{eqnarray}
\langle \text{T}_{c} X_m(\tau)X_{m'}^\dagger(\tau')
\rangle_{\bar{H}}=\text{exp}\left(\sum_{\nu,\nu'}i\frac{\lambda_{\nu, m}\lambda_{\nu',m'}}{\Omega_\nu
\Omega_{\nu'}}D_{\nu,\nu'}(\tau,\tau')-i\frac{\lambda_{\nu, m}^2+\lambda_{\nu',m'}^2}{2\Omega_\nu
\Omega_{\nu'}}D_{\nu,\nu'}(\tau,\tau)\right),
\end{eqnarray}
where we use the momentum correlation functions 
\begin{eqnarray}
D_{\nu,\nu'}=-i\langle \text{T}_{c} p_\nu(\tau)p_{\nu'}(\tau')\rangle_{\bar{H}}.
\end{eqnarray}
Employing the equation of motion for $D_{\nu\nu'}$
\begin{eqnarray}
\frac{1}{4\Omega_{\nu}\Omega_{\nu'}} (-\partial^{2}_{\tau}-\Omega_{\nu}^{2}) D_{\nu,\nu'}(\tau,\tau') (-\partial^{2}_{\tau'}-\Omega_{\nu'}^{2}) &=& \delta(\tau,\tau') (-\partial^{2}_{\tau'}-\Omega_{\nu'}^{2}) \frac{1}{2\Omega_{\nu'}} \\ 
&&+ \Pi_{\text{bath},\nu,\nu'}(\tau,\tau')+ \Pi_{\text{el},\nu,\nu'}(\tau,\tau'), \nonumber
\end{eqnarray}
we determine the corresponding self-energy matrices $\Pi_{\text{bath},\nu,\nu'}$ and $\Pi_{\text{el},\nu,\nu'}$. The self-energy matrix 
\begin{eqnarray}
\Pi_{\text{bath},\nu,\nu'} &=& \delta_{\nu,\nu'} \sum_{\beta_{c}} \vert \eta_{\nu,\beta_{\nu}} \vert^{2} D^{0}_{\beta_{\nu}}(\tau,\tau')
\end{eqnarray}
includes the coupling of the vibrational modes to the thermal baths, where $D^{0}_{\beta_{\nu}}$ denotes the free Green's function 
of bath-mode $\beta_{\nu}$. 
The electronic self-energy part, $\Pi_{\text{el},\nu,\nu'}$, describing 
the interactions between the vibrational modes and 
the electronic degrees of freedom of the molecular bridge, 
is evaluated to second order in the molecule-lead coupling \cite{Galperin06,Hartle,MSVE10}
\begin{eqnarray}
\label{Piel}
\Pi_{\text{el},\nu,\nu'}(\tau,\tau')=-i\sum_{m,m'}\frac{\lambda_{\nu, m}\lambda_{\nu',m'}}{\Omega_\nu \Omega_{\nu'}}(\Sigma_{m,m'}(\tau,\tau')\bar{G}_{m',m}(\tau',\tau)+\Sigma_{m',m}(\tau',\tau)\bar{G}_{m,m'}(\tau,\tau')).
\end{eqnarray}
Thereby, we use the noncrossing approximation, where contributions mixing mode-bath and molecule-lead couplings are disregarded.
Since $\Pi_{\text{el},\nu,\nu'}$ depends on the electronic self-energies $\Sigma_{m,m'}=\Sigma_{\text{L},m,m'}+\Sigma_{\text{R},m,m'}$ and Green's functions $\bar{G}_{m,m'}$, the respective Dyson-Keldysh equations need to be solved 
iteratively in a self-consistent scheme \cite{Galperin06,Hartle}.

With these Green's function, $D_{\nu,\nu'}$ and $G_{m,m'}$, 
the vibrational excitation of each vibrational mode is obtained according to \cite{Hartle,MSVE10}:
\begin{eqnarray}
\label{formulavibex}
\langle c^\dagger_\nu c_\nu \rangle_{\hat{H}}&\approx&-\left(A_{\nu}+\frac{1}{2}\right)\text{Im}\left[D^{<}_{\nu,\nu}(t=0)\right] -\left(B_{\nu}+\frac{1}{2}\right)\\
&&+\sum_{m}\frac{\lambda_{\nu, m}^{2}}{\Omega_\nu^2}\text{Im}[\bar{G}^<_{m,m}(t=0)] \nonumber\\
&&+2\sum_{m<m'}\frac{\lambda_{\nu, m}\lambda_{\nu, m'}}{\Omega_\nu^2}\text{Im}[\bar{G}^<_{m,m}(t=0)]\text{Im}[\bar{G}^<_{m',m'}(t=0)] \nonumber\\
&&-2\sum_{m<m'}\frac{\lambda_{\nu, m}\lambda_{\nu, m'}}{\Omega_\nu^2}\text{Im}[\bar{G}^<_{m',m}(t=0)]\text{Im}[\bar{G}^<_{m',m}(t=0)], \nonumber\\
A_{\nu}&=&\sum_{\beta_{\nu}} \mathcal{P}\frac{\vert \eta_{\nu,\beta_{\nu}}\vert^{2}\omega_{\beta_{\nu}}}{\Omega_{\nu}(\omega_{\beta_{\nu}}^{2}-\Omega_{\nu}^{2})},\quad  \\ 
B_{\nu}&=&\sum_{\beta_{\nu}} \mathcal{P}\frac{\vert \eta_{\nu,\beta_{\nu}}\vert^{2}}{\omega_{\beta_{\nu}}^{2}-\Omega_{\nu}^{2}}\left(1+2n_{\text{u}}(\omega_{\beta_{\nu}})\right),
\end{eqnarray}
which is consistent with the second order expansion used for 
the evaluation of the associated Green's functions. 
In Eq.\ (\ref{formulavibex}), we also use the Hartree-Fock factorization: $\langle a_{m}^{\dagger}a_{m} a_{m'}^{\dagger}a_{m'} \rangle\approx\langle a_{m}^{\dagger}a_{m} \rangle\langle a_{m'}^{\dagger}a_{m'} \rangle-\langle a_{m}^{\dagger}a_{m'} \rangle\langle a_{m'}^{\dagger}a_{m} \rangle$.
The current is calculated employing the Meir-Wingreen formula \cite{Meir92}
\begin{eqnarray}
I &=& 2e\int\frac{\text{d}\epsilon}{2\pi}\, \sum_{m,m'} \left( \Sigma_{\text{L},m,m'}^{<}(\epsilon)\bar{G}^{>}_{m',m}(\epsilon)-\Sigma_{\text{L},m,m'}^{>}(\epsilon)\bar{G}^{<}_{m',m}(\epsilon) \right) .
\end{eqnarray}
Note that this scheme, including the elastic co-tunneling approximation, is current-conserving.

\section{Results}
\label{results}

In this section, we employ 
the ME and the NEGF methodology, outlined in Secs.\ \ref{MEapproach} 
and \ref{NEGFapproach}, respectively, 
to analyze transport characteristics of asymmetric 
single-molecule junctions. 
In particular, we consider a 
single vibrational mode coupled 
to a single electronic state in Sec.\ \ref{VSE}, 
and demonstrate that an external bias voltage can 
be used to control its level of excitation. 
In Sec.\ \ref{MSVE}, we extend these studies to two 
vibrational modes  
and show that their level of excitation, despite their different frequencies, 
can be selectively controlled by switching the polarity of the applied bias voltage. 
To demonstrate the generality of the phenomenon and to 
corroborate our findings, we consider a variety of parameter regimes,  
including the effect of intra-molecular interactions.

\subsection{Bias-Controlled Excitation of a Single Vibrational Mode in an Asymmetric Molecular Junction}
\label{VSE}

We first consider a model with a 
single vibrational mode coupled 
to a single electronic state in Sec.\ \ref{VSE} and study how the excitation 
of the vibrational mode 
can be controlled by an external bias voltage $\Phi$. 
To this end, we consider a vibrational mode with frequency $\Omega=0.15$\,\text{eV} 
that is coupled to a single electronic state with coupling strength $\lambda_{1,1}=0.6\Omega_{1}$.  
The electronic state is located $\epsilon_{1}=0.6$\,eV above the Fermi-level, and  
is asymmetrically coupled to the left, $\upsilon_{\text{L},1}=0.1$, 
and to the right lead, $\upsilon_{\text{R},1}=0.03$, respectively. 
The specific model parameters are detailed in Table \ref{tableI}. 
These parameters (including those given in Tab.\ \ref{tableII}) reflect typical values for 
molecular junctions, as they are determined for example in ab-initio calculations  \cite{Pecchia04,Frederiksen04,Benesch06,Troisi2006,Benesch06,Benesch08,Benesch09,Monturet2010} or experiments \cite{WuNazinHo04,Natelson2004,Huang2006,Natelson2008,Repp,Ballmann2010,Tao2010,Osorio2010,Secker2011}.

\begin{center}
\begin{table}
\caption{\label{Table I}Model Parameters (Energy values are given in eV, $K\in\{\text{L,R}\}$)}
\begin{center}
\begin{tabular}{|*{12}{c|}}
\hline \hline
 & $\epsilon_{1}$ &  $\upsilon_{\text{L},1}$ & $\upsilon_{\text{R},1}$ & \;$\xi$\; & \;$\gamma$\; & $\Gamma_{K}(\mu_{K})$ & $\omega_{c,1}$ & \;\;$k_{\text{B}}T$\;\; & \;\;$\Omega_{1}$\;\; & \;$\lambda_{1,1}$\; & $\zeta_{1}$ \\
\hline
Figs.\ \ref{Fig1a} and \ref{VSE2} & 0.6 & 0.1 & 0.03 & 1 & 2 & 0.01 & 1 & 0.001 & 0.15 & 0.09 & 0 \\
Fig.\ \ref{varye0} & 0 -- 1 & 0.1 & 0.03 & 1 & 2 & 0.01 & 1 & 0.001 & 0.15 & 0.09 & 0 \\ 
Fig.\ \ref{varyratio} & 0.6 & 0.1 & 0.01 -- 0.1 & 1 & 2 & 0.01 & 1 & 0.001 & 0.15 & 0.09 & 0 \\
Fig.\ \ref{Fig.7} & 0.6 & 0.1 & 0.03 & 1 & 2 & 0.01 & 1 & 0.001 & 0.15 & 0.09 & 0 -- 0.04\\
\hline \hline
\end{tabular}
\end{center}
\label{tableI}
\end{table}
\end{center}

\begin{figure} 
\begin{center}
\resizebox{\newwidth}{\newheight}{
\includegraphics{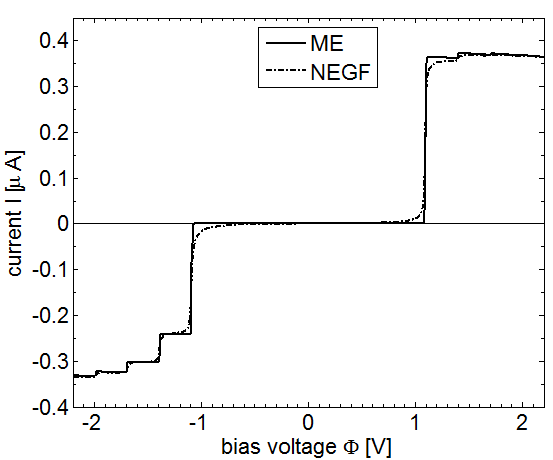}
}
\resizebox{\newwidth}{\newheight}{
\includegraphics{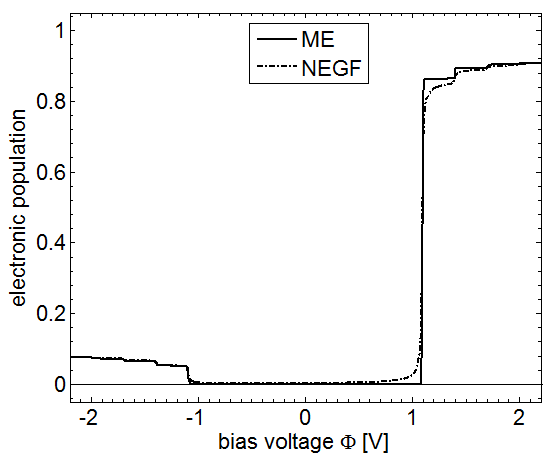}
}
\end{center}
  \caption{\emph{Upper Panel:} Current-voltage characteristics for a model molecular junction comprising a single  electronic state that is coupled to a single vibrational mode and asymmetrically to a left and a right lead. The  solid black line is obtained with the ME approach, while for the dashed black line NEGF is used. \emph{Lower Panel:} The corresponding population of the electronic state as a function of the applied bias voltage $\Phi$. The asymmetry of the population characteristics with respect to the polarity of the bias voltage is a result of the asymmetric molecule-lead coupling. Electronic-vibrational coupling translates this asymmetry also to the respective current-voltage characteristics, which otherwise (\emph{i.e.}\ without electronic-vibrational coupling) would be almost anti-symmetric with respect to $\Phi$.}\label{Fig1a}
\end{figure}

\begin{figure}
\begin{center}
\begin{tabular}{cccc}
\resizebox{\newwidthprime}{\newheightprime}{
\includegraphics{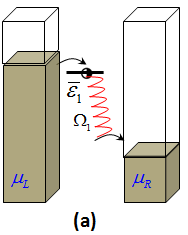}
}
&
\resizebox{\newwidthprime}{\newheightprime}{
\includegraphics{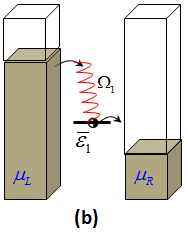}
}
&
\resizebox{\newwidthprime}{\newheightprime}{
\includegraphics{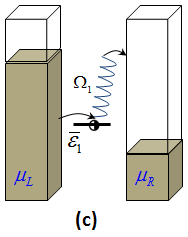}
}
&
\resizebox{\newwidthprime}{\newheightprime}{
\includegraphics{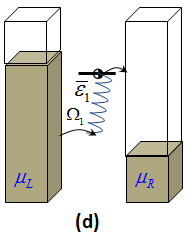}
}
\\ 
\end{tabular}
\end{center}
\caption{(Color online) \label{basmech} Basic schemes of vibrationally 
coupled electron transport through a single electronic state. 
Panels a) and b) depict examples for emission processes, where an electron 
sequentially tunnels from the left lead onto the molecule and further to the right lead, 
thereby singly exciting the vibrational mode of the molecular bridge (red wiggly lines). 
Such emission processes are effectively 'heating' the junction (local heating). 
Examples for respective absorption processes are shown in Panel c) and d), where  
electrons tunnel from the left to the right lead by absorbing a quantum of vibrational 
energy (blue wiggly line). 
}
\end{figure}

\begin{figure} 
\begin{center}
\resizebox{\newwidth}{\newheight}{
\includegraphics{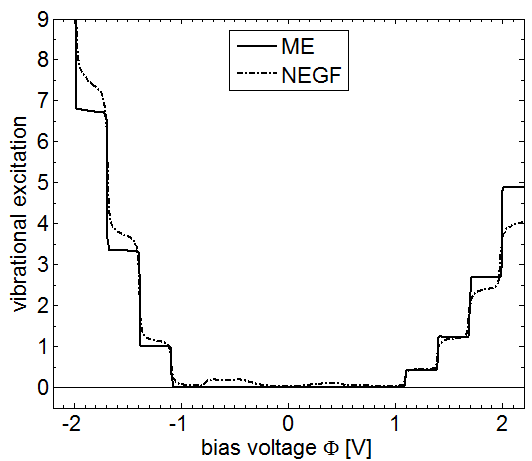}
}
\end{center}
  \caption{(Color online) Vibrational excitation characteristics corresponding to the 
current-voltage and population characteristics shown in Fig.\ \ref{Fig1a}. 
Due to the asymmetry in the 
molecule-lead coupling of this model molecular junction, 
electron-hole pair creation processes are cooling the vibrational mode 
more efficiently for positive bias voltages 
than for negative ones, leading to a strongly asymmetric excitation characteristics. 
The external bias voltage thus can be used to control the 
level of vibrational excitation in this asymmetric molecular junction.}\label{VSE2}
\end{figure}

Current-voltage characteristics for this model molecular junction 
and the respective population of the electronic state are shown in Fig.\ \ref{Fig1a}. 
Thereby, the solid black lines represent results obtained with the 
reduced density matrix approach, while the dashed black lines show results, 
for which we employed the nonequilibrium Green's function approach. 
The results of both approaches agree very well.  
Minor deviations between the 
approaches occur due to 
the broadening of the molecular levels, 
which is not included in the ME scheme. 
For positive bias voltages, the current and the population of the electronic state  
display a single step. This step indicates the onset of 
transport at $e\Phi=2\bar{\epsilon}_{1}$ 
($\bar{\epsilon}_{1}={\epsilon}_{1}-{\lambda}_{1,1}^{2}/{\Omega_1}$ denotes  
the polaron-shifted energy of state 1 (cf.\ Sec.\ \ref{NEGFapproach})), 
where electrons from the left lead 
can resonantly tunnel onto the molecular bridge. 
Notice that the molecular energy level is located well above 
the Fermi-level of the junction, that is  
by several units of the vibrational frequency: 
$\bar{\epsilon}_{1}-\epsilon_{\text{F}}>3\Omega_{1}$. 
Therefore, at the onset of transport by electron tunneling from the left electrode, 
several inelastic channels corresponding to 
processes described in Figs.\ \ref{basmech}a, \ref{basmech}c and 
\ref{basmech}d open up 
simultaneously. As the bias increases further, additional heating channels become available, 
involving tunneling of high energy electrons from the strongly coupled (left) electrode 
onto the molecule (see Fig.\ \ref{basmech}b). However, 
these additional channels do not significantly increase the current, 
since the bottleneck for transport in this asymmetric junction is 
tunneling processes from the molecular bridge to the weakly coupled (right) 
electrode that are already active. Accordingly, the electronic state is populated 
(from the left) much faster than it is depopulated by tunneling processes 
to the right, and is therefore almost fully occupied for $e\Phi>2\bar{\epsilon}_{1}$. 
For negative bias voltages, however, both the current and the electronic population 
exhibit a number of pronounced steps at 
$\Phi=-2(\bar{\epsilon}_{1}+n\Omega_{1})$ ($n\in\mathbb{N}_{0}$). Again, 
different inelastic transport channels open up simultaneously 
at the onset of the current. However, in this case, 
as the bias decreases further, $\Phi<-2\bar{\epsilon}_{1}$, 
additional heating channels open up one by one at the bottleneck for transport, 
that is additional tunneling processes with respect to the right lead. 
Since these processes are inactive for higher negative bias voltages, 
$-2\bar{\epsilon}_{1}<\Phi<0$, 
one observes significant steps in the current-voltage as well as in the respective 
population characteristics. 
Notice that in this bias direction the electronic state is depopulated (to the left) 
much faster than it is populated (from the right), so that it is almost unoccupied for
 $e\Phi<-2\bar{\epsilon}_{1}$. 
The relative step heights that occur in these characteristics 
qualitatively reflect the transition probabilities  $\frac{1}{n!}\left(\frac{\lambda_{1,1}}{\Omega_{1}}\right)^{2n}\text{e}^{-(\lambda_{1,1}/\Omega_{1})^{2}}$ 
for a transition from the vibrational ground- to its $n$th excited state. 
For a quantitative analysis of the step heights, however, 
the nonequilibrium state of the vibrational mode, 
which is typically highly excited, needs to be considered (cf.\ Fig.\ \ref{VSE2}). 
As a result of electronic-vibrational coupling and the asymmetry in the coupling to the leads, 
the current-voltage characteristics thus 
exhibits a pronounced asymmetry with respect to the polarity of the applied bias voltage $\Phi$, 
which is also referred to as vibrational rectification. 
This has been theoretically analyzed 
\cite{Flensberg03,Hartle2010b} and experimentally 
verified \cite{WuNazinHo04,Schulze08,Ballmann2010} before.

\begin{figure}
\begin{center}
\begin{tabular}{c}
\resizebox{\newwidthprime}{\newheightprime}{
\includegraphics{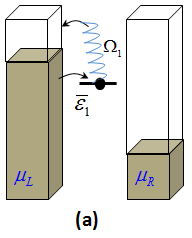}
}
\resizebox{\newwidthprime}{\newheightprime}{
\includegraphics{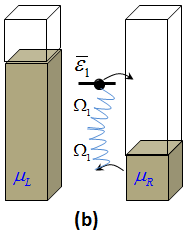}
}
\resizebox{\newwidthprime}{\newheightprime}{
\includegraphics{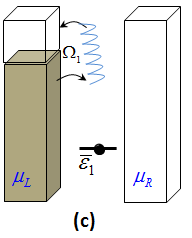}
}
\\ 
\end{tabular}
\end{center}
\caption{(Color online) \label{el-h-pair-creation} Example electron-hole pair 
creation processes in a molecular junction. 
Panel a) depicts an electron-hole pair creation process with respect 
to the left lead by absorption of a single vibrational quantum. 
Panel b) represents an electron-hole pair 
creation process with respect to the right lead by absorption of two vibrational quanta. 
The absorption of two vibrational quanta typically occurs with lower probability. 
Panel c) shows an off-resonant electron-hole pair creation process.}
\end{figure}

In contrast to these electronic observables, 
the corresponding vibrational excitation number, 
$\langle c^\dagger_\nu c_\nu \rangle_{\hat{H}}$, 
increases in a series of 
distinct steps for both polarities of the bias voltage (see Fig.\ \ref{VSE2}). 
This finding cannot be solely understood in terms of electron transport processes. 
Although vibrational excitation is a result of inelastic 
electron transport processes (cf.\ Figs.\ \ref{basmech}a-d), 
another class of processes, which does not contribute to the current, 
needs to be considered. Resonant electron-hole pair creation processes 
\cite{MSVE10,Hartle2010b,Hartle2011}, such as those depicted by 
Figs.\ \ref{el-h-pair-creation}a and \ref{el-h-pair-creation}b, are effectively 
cooling the vibrational mode and diminish the current-induced vibrational excitation. 
Since these processes involve two sequential tunneling events, they occur with 
the same probability as respective transport processes. 
Due to the asymmetry in the molecule-lead coupling, 
electron-hole pair creation processes 
with respect to the left lead are the most important ones. 
They are typically more effective the less vibrational quanta are involved. 
For this particular model system, 
cooling by electron-hole pair creation is therefore more 
pronounced for positive bias voltages, $\Phi>2\bar{\epsilon}_{1}$, 
where \emph{e.g.}\ an electron-hole pair in the left 
lead can be produced by absorption of just a single quantum of vibrational energy. 
For negative bias voltages, $\Phi<-2\bar{\epsilon}_{1}$, the creation of an 
electron-hole pair in the left lead requires the absorption of more than 
ten vibrational quanta. As a result, vibrational excitation is much smaller 
for positive bias voltages than for negative ones. 
Increasing the bias voltage, these pair creation processes are blocked 
one by one, 
as the energy gap between the molecular level and the electrode 
chemical potential increases. 
The steps in the vibrational excitation characteristics 
thus become larger with increasing bias voltage due to less efficient 
cooling by electron-hole pair creation processes \cite{Hartle2010b,Hartle2011}. 
This blocking of pair creation processes appears 
for both polarities of the bias voltage. 

Apart from the broadening of steps, the ME and the NEGF approach give almost the 
same vibrational excitation characteristics.   
However, NEGF can be expected to give slightly larger values 
for the vibrational excitation, because inelastic co-tunneling processes 
\cite{Lueffe,Hartle,Hartle09,Leijnse09}, which are not included in the ME scheme, 
additionally contribute to the level of excitation for the vibrational mode. 
This is particularly important in the off-resonant transport regime, \emph{i.e.}\ for $\vert\Phi\vert<\bar{\epsilon}_{1}$, where NEGF gives a small vibrational excitation 
while ME does not. 
Significant deviations between both approaches occur only for large bias voltages. 
Especially for large positive bias voltages, \emph{e.g.}\ at $\Phi>2$\,V, 
the vibrational excitation obtained from NEGF is significantly smaller than 
the one obtained by the ME scheme. 
We attribute this behavior to 
the contribution of cooling by 
off-resonant electron-hole pair creation processes,
which are missing in the ME approach 
(an example of an off-resonant pair creation process 
is depicted in Fig.\ \ref{el-h-pair-creation}c). 
These processes become the dominant cooling mechanism 
at large bias voltages, 
where resonant electron-hole pair creation processes are suppressed, as they 
require increasingly higher vibrational energy. 
In contrast, off-resonant pair creation processes can occur 
by the absorption of 
just a single quantum of vibrational energy for all bias voltages. 
The ME approach thus gives a somewhat larger vibrational excitation 
than the NEGF method for bias voltages, where resonant electron-hole pair creation processes 
are strongly suppressed.

\begin{figure} 
\begin{center}
\resizebox{\newwidth}{\newheight}{
\includegraphics{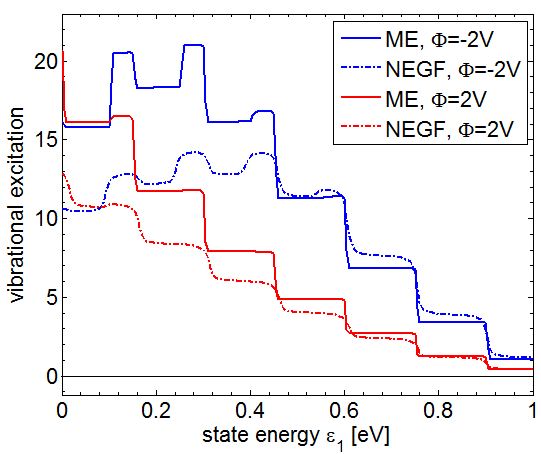}
}
\resizebox{\newwidth}{\newheight}{
\includegraphics{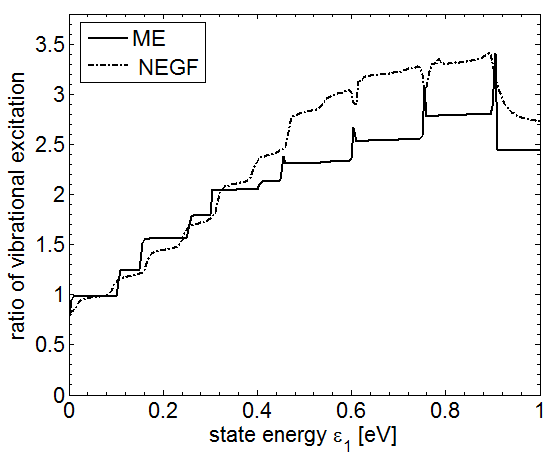}
}
\end{center}
\caption{(Color online) \emph{Upper Panel}: Vibrational excitation as a function of the energy, $\epsilon_{1}$, 
for a molecular junction with an electronic state coupled to a single vibrational mode and 
asymmetrically to the leads. Red and blue lines refer to results that are obtained for a fixed bias voltage, $\Phi=\pm2$\,V, respectively. \emph{Lower Panel}: Ratio of the average vibrational 
excitation numbers $\langle c_{1}^{\dagger}c_{1}\rangle_{\Phi=-2\text{\,V}}/\langle c_{1}^{\dagger}c_{1}\rangle_{\Phi=+2\text{\,V}}$ shown in the upper panel. 
The further the electronic level is located from the Fermi-level of the system, which can 
be controlled for example by a gate electrode, the more pronounced is the 
asymmetry in vibrational excitation for the different 
polarities of the bias voltage $\Phi$.}\label{varye0}
\end{figure}

The importance of cooling by electron-hole pair creation processes 
can be corroborated by studying the behavior of 
the vibrational excitation characteristics with respect 
to the energy of the electronic state, which for a given bias voltage 
influences the efficiency of resonant electron-hole pair creation processes. 
Fig.\ \ref{varye0}a shows results for the vibrational excitation as a function 
of the energy $\epsilon_{1}$ at a fixed bias voltage. Thereby, the blue lines 
refer to a fixed bias voltage of $\Phi=-2$\,V, and the red lines to $\Phi=2$\,V. 
As before, solid (dashed) lines refer to calculations performed with the ME (NEGF) scheme. 
If the energy of the electronic state is closer to the Fermi-level of the system, 
$\bar{\epsilon}_{1} \to 0$, resonant electron-hole pair creation processes are 
more strongly suppressed, as they require the absorption of 
an increasing number of vibrational quanta. 
The less efficient cooling by resonant electron-hole pair creation 
leads to the general observed trend of    
an increasing vibrational excitation with a decreasing energy 
of the electronic state (for example from $\epsilon_{1}=1$\,eV to $\epsilon_{1}=0.4$\,eV). 
For $\epsilon_{1}<0.4$\,eV, 
the NEGF scheme gives a significantly smaller vibrational excitation than the ME method. 
Here, cooling by off-resonant electron-hole pair creation processes 
(missing in the ME treatment) 
results in a significantly lower level of vibrational excitation. 
Notice that for an asymmetric junction, the value of $\bar{\epsilon}_{1}$ for which 
the vibrational excitation obtains its maximal value differs from zero and depends on 
the bias polarity. Considering, \emph{e.g.}\ the negative bias voltage (blue lines), 
a maximum of vibrational excitation is obtained at $\bar{\epsilon_{1}}\sim0.3$\,eV ($\epsilon_{1}\sim0.35$\,eV). 
For this bias, shifting the electronic level to lower values 
enhances cooling at the left electrode and suppresses cooling at the right electrode. 
Due to the asymmetry in the molecule-lead coupling, pair creation processes 
with respect to the left lead are more important than with respect to the right lead.   
A minimum of cooling efficiency by electron-hole pair creation processes, which  
corresponds to a maximum in vibrational excitation, is thus reached for  
positive values of $\bar{\epsilon}_{1}$. 
Similarly, for positive bias, a maximum in vibrational excitation appears for 
negative values of $\bar{\epsilon}_{1}$. 
Fig.\ \ref{varye0}b represents the ratio $\langle c_{1}^{\dagger}c_{1}\rangle_{\Phi=-2\text{\,V}}/\langle c_{1}^{\dagger}c_{1}\rangle_{\Phi=+2\text{\,V}}$. It shows that the asymmetry in the vibrational excitation characteristics, 
as well as in 
the current-voltage characteristics and the electronic population (data not shown), 
disappears, once the electronic level is located close to the Fermi-level of the system. 
This demonstrates that for an electronic state close to the Fermi-level, 
which can be controlled for example by a gate electrode 
\cite{Sapmaz05,Song2009,Osorio2010,Martin2010},  
the efficiency of both current-induced heating and  
cooling by electron-hole pair creation processes 
is the same for both polarities of the bias voltage $\Phi$.

\begin{figure} 
\begin{center}
\resizebox{\newwidth}{\newheight}{
  \includegraphics{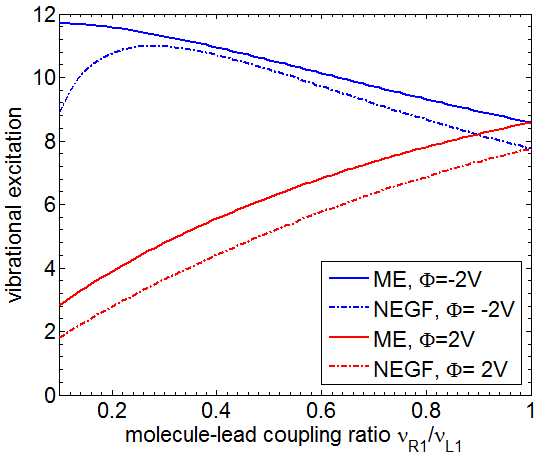}
}
\end{center}
  \caption{(Color online) \label{varyratio} Vibrational excitation of a molecular junction with a single electronic state coupled to a single vibrational mode for different coupling strengths to the right lead, $\upsilon_{\text{R},1}$. Thereby, the coupling to the left lead is fixed,  $\upsilon_{\text{L},1}=0.1$\,eV, as is the bias voltage, $\Phi=\pm2$\,V, for the red and the blue lines, respectively. 
Asymmetric coupling to the leads facilitates the control of vibrational excitation 
by the polarity of an external bias voltage.}
\end{figure}

At this point,  
it is interesting 
to study 
the extent of  
vibrational excitation 
for different ratios of the molecule-lead couplings 
$\upsilon_{\text{R},1}/\upsilon_{\text{L},1}$. In Fig.\ \ref{varyratio}, 
we show the level of excitation of the vibrational mode as a function of the 
ratio $\upsilon_{\text{R},1}/\upsilon_{\text{L},1}$, where 
$\upsilon_{\text{L},1}=0.1$\,eV is fixed. Again, red and blue  
lines refer to calculations performed at a fixed bias voltage of $\Phi=\pm 2$\,V, respectively. 
Trivially, for a symmetric junction with 
$\upsilon_{\text{R},1}=\upsilon_{\text{L},1}$, we obtain the same level of 
vibrational excitation for both polarities of the bias voltage. 
Decreasing the coupling to the right lead, 
the asymmetry in vibrational excitation increases almost linearly. 
Interestingly, for negative bias voltages, the vibrational excitation 
obtained by the NEGF scheme (dashed blue line) and by the ME method 
(solid blue line) approach one another upon decreasing $\upsilon_{\text{R},1}$. 
This points to the fact that off-resonant electron-hole pair creation 
processes with respect to the right lead become strongly suppressed. 
For even smaller coupling strengths to the right lead, 
$\upsilon_{\text{R},1}<0.03$\,eV, the turnover in the dashed blue line (NEGF scheme) 
indicates that in the limit $\upsilon_{\text{R},1}\rightarrow0$ 
current-induced vibrational excitation vanishes, as does the corresponding current. 
The solid blue line (ME scheme) exhibits the same turn-over 
but for even smaller coupling strengths to the right lead. 
On the other hand, for $\Phi=2$\,V, the dashed and solid red lines 
remain well separated upon decreasing the coupling to the right lead, 
as the ratio between resonant and off-resonant pair creation processes 
with respect to the left lead remains constant.

\begin{figure} 
\begin{center}
\resizebox{\newwidth}{\newheight}{
\includegraphics{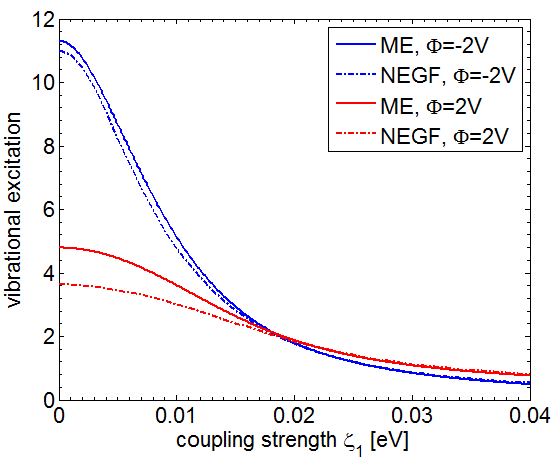}
}
\end{center}
  \caption{(Color online) Vibrational excitation as a function of the mode-bath coupling strength, $\zeta_{1}$, 
for a molecular junction with an electronic state coupled to a single vibrational mode and 
asymmetrically to the leads. Red and blue lines refer to results that are obtained for a fixed bias voltage, $\Phi=\pm2$\,V, respectively. For a strong coupling between the vibrational mode and the thermal bath, vibrational  excitation is governed by the formation of a polaronic state.}\label{Fig.7}
\end{figure}

So far, we have discussed cooling mechanisms 
for the vibrational mode, which are solely induced by electronic-vibrational coupling. 
Other vibrational energy relaxation processes, which can be of relevance in molecular junctions, 
include Intramolecular Vibrational Energy Redistribution processes (IVR) or energy 
transfer to the environment (\emph{e.g.}\ phononic excitation of the electrodes) \cite{Segal2000,Lehmann04,May06,Tao2006,Seidemann10}. 
Such relaxation mechanisms are commonly described by coupling of the primary 
vibrational mode(s) to a thermal heat bath. 
Fig.\ \ref{Fig.7} shows the level of vibrational excitation 
as a function 
of the mode-bath coupling strength $\zeta_{1}$ for a fixed bias voltage 
(blue lines correspond to $\Phi=-2$\,V, and red lines to $\Phi=+2$\,V). 
Naturally, the molecular junction responds to an increased coupling to a 
"cold" thermal bath by decreased levels of current-induced vibrational excitation. 
However, as can be seen by inspection of 
Eqs.\ (\ref{formulavibex}), vibrational excitation 
is not only a result of 
inelastic 
transport processes, but also stems from the population of the electronic states, that is  
the formation of a polaronic state \cite{Thoss2011}. 
Such polaron-formation
leads to a finite vibrational excitation 
even in the limit of strong mode-bath coupling $\zeta_{1}$. 
Since the electronic level is almost fully populated for positive, but almost 
unoccupied for negative bias voltages, one observes a higher 
vibrational excitation for positive bias voltages than for negative bias voltages, if 
the mode-bath coupling strength $\zeta_{1}$ exceeds a value of $0.02$\,eV.

We finally conclude that in an asymmetric molecular junction the level of vibrational excitation 
can be controlled by the magnitude and the polarity of 
the applied bias voltage. 
It is noted that a gate voltage, which allows to align the energy of electronic states, $\epsilon_{i}$, 
with respect to the Fermi-level, may facilitate a control mechanism for the ratio between 
the different levels of vibrational excitation at different bias polarities.

\subsection{Mode-Selective Vibrational Excitation}
\label{MSVE}

\begin{figure} 
\begin{center}
\includegraphics[scale=0.75]{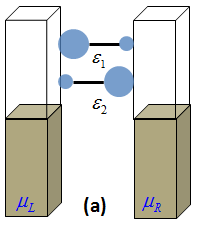}
\includegraphics[scale=0.75]{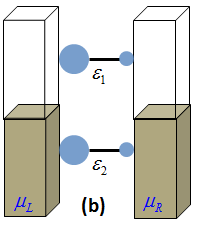}
\end{center}
  \caption{(Color online) Two generic model systems for a molecular junction exhibiting MSVE. 
Panel (a) depicts a molecular junction, where two electronic states are located above 
the Fermi-level of the junction, while panel (b) shows a molecular junction that 
involves an electronic state above and another state below the Fermi-level. 
Strong (weak) coupling of the electronic states to the leads is represented 
by large (small) blue dots.}\label{Fig.8} 
\end{figure}

In Sec.\ \ref{VSE} we have outlined how the level of 
excitation of a single vibrational mode can be controlled by an external 
bias voltage $\Phi$. 
In this section, we extend this concept 
to selective excitation of specific vibrational modes in a junction 
with multiple vibrational degrees of freedom. 
In particular, we show that modes with higher frequencies 
can be stronger excited than low-frequency modes, 
which corresponds to a "non-statistical" distribution of vibrational energy. 
A minimal model for two vibrational modes (model A), which demonstrates such 
mode-selective vibrational excitation, 
was recently introduced \cite{MSVE10}. It involves two electronic states, 
where each state is coupled to one of the vibrational modes and 
asymmetrically to the leads.  
Thereby, the asymmetry in the coupling to the leads reflects  
an inherent asymmetry of the contacted molecule. 
In this section, we review and extend 
our earlier study of model A, 
taking into account  
intra-molecular correlations, in particular off-diagonal electronic-vibrational 
coupling, $\lambda_{\nu,m}\neq\delta_{\nu m}$, 
and electron-electron interactions, $U_{m,n}\neq0$. 
Moreover, we consider a different generic realization of an 
asymmetric molecular junction exhibiting MSVE, model B. 
Model B also comprises two vibrational modes and two electronic states, 
asymmetrically coupled to leads, 
but in contrast to model A, the asymmetry in the molecule-lead coupling 
is not a result of an 
intrinsic asymmetry of the molecule, but rather stems 
from an asymmetry in the electrodes. 
This corresponds for example to an STM setup, where 
the molecule bridging the gap between the two electrodes 
is typically much stronger coupled to the substrate 
than to the STM tip. 
These two scenarios, where MSVE can be controlled by an external bias voltage, 
are schematically depicted in Fig.\ \ref{Fig.8}. 
Respective model parameters are detailed in Table \ref{tableII}. 
Note that an asymmetric molecule-lead as well as electronic-vibrational coupling 
is necessary to observe MSVE in both model systems 
(cf.\ the discussion of Fig.\ \ref{varyratio}). 

\begin{table}
\caption{\label{Table II}Model Parameters (Energy values are given in eV, $K\in\{\text{L,R}\}$, $\alpha\in\{\text{1,2}\}$)}
\begin{center}
\begin{tabular}{|*{14}{c|}}
\hline \hline
 & $\epsilon_1,\epsilon_2$ &  $\upsilon_{\text{L},1},\upsilon_{\text{L},2}$ & $\upsilon_{\text{R},1},\upsilon_{\text{R},2}$ & $\xi$ & $\gamma$ & $\Gamma_{K}(\mu_{K})$ & $\omega_{c,\alpha}$ & $k_{\text{B}}T$ & $\Omega_1, \Omega_2$ & $\lambda_{1,1},\lambda_{2,2}$ & $\alpha$ & $U_{1,2}$ & $\zeta_{\alpha}$ \\
\hline
Fig.\ \ref{Fig.9} & 0.65,0.575 & 0.1,0.03 & 0.03,0.1 & 1 & 2 & 0.01 & 1 & 0.001 & 0.15,0.2 &0.09,0.12&0 & 0 & 0\\
Fig.\ \ref{Fig.9c} & 0.65,0.575 & 0.1,0.03 & 0.03,0.1 & 1 & 2 & 0.01 & 1 & 0.001 & 0.15,0.2 &0.09,0.12&0 & 0 & 0\\
Fig.\ \ref{Fig.10} & 0.65,-0.5 & 0.1,0.1 & 0.03,0.03 & 1 & 2 & 0.01 & 1 & 0.001 & 0.15,0.2 &0.09,0.12&0 & 0 & 0\\
Fig.\ \ref{Fig.10c} & 0.65,-0.5 & 0.1,0.1 & 0.03,0.03 & 1 & 2 & 0.01 & 1 & 0.001 & 0.15,0.2 &0.09,0.12&0 & 0 & 0\\
Fig.\ \ref{Fig.11} & 0.65,0.575 & 0.1,0.03 & 0.03,0.1 & 1 & 2 & 0.01 & 1 & 0.001 & 0.15,0.2 &0.09,0.12& 0 -- 1 & 0 & 0\\
Fig.\ \ref{Fig.12} & 0.65,-0.5 & 0.1,0.1 & 0.03,0.03 & 1 & 2 & 0.01 & 1 & 0.001 & 0.15,0.2 &0.09,0.12& 0 -- 1 & 0 & 0\\
Fig.\ \ref{Fig.15} & 0.65,0.575 & 0.1,0.03 & 0.03,0.1 & 1 & 2 & 0.01 & 1 & 0.001 & 0.15,0.2 &0.09,0.12&0 & 0 -- 1.5 & 0\\
Fig.\ \ref{Fig.16} & 0.65,-0.5 & 0.1,0.1 & 0.03,0.03 & 1 & 2 & 0.01 & 1 & 0.001 & 0.15,0.2 &0.09,0.12&0 & 0 -- 1.5 & 0\\
\hline \hline
\end{tabular}
\end{center}
\label{tableII}
\end{table}

\subsubsection{The Basic Phenomenon}

\begin{figure}
\begin{center}
\resizebox{\newwidth}{\newheight}{
\includegraphics{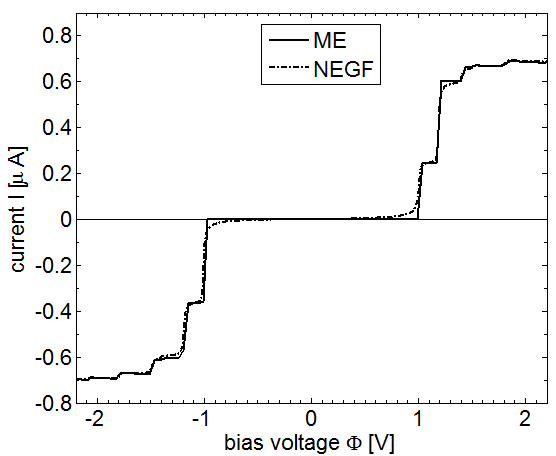}
}
\resizebox{\newwidth}{\newheight}{
\includegraphics{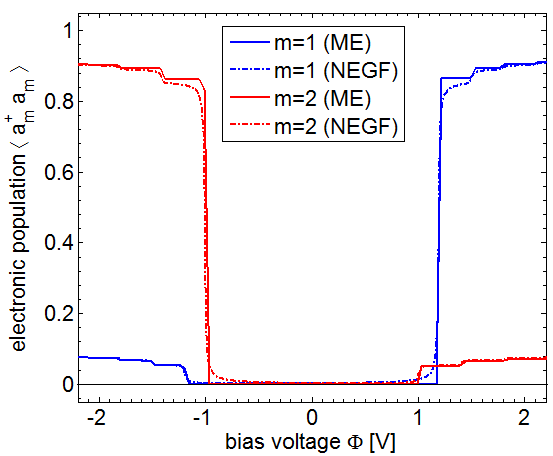}
}
\end{center}
  \caption{(Color online) Current-voltage characteristics and the respective population of the electronic states for the model system depicted by Fig.\ \ref{Fig.8}a (model A). Solid (dashed dotted) lines are obtained employing the ME (NEGF) methodology. The population characteristics of state 1 and 2 are depicted by the blue and the red line,  respectively. While the current is almost anti-symmetric with respect to the bias voltage $\Phi$, the population of the electronic levels reflects the asymmetry in the coupling of the two states to the left and the right lead.}\label{Fig.9}
\end{figure}

\begin{figure}
\begin{center}
\resizebox{\newwidth}{\newheight}{
\includegraphics{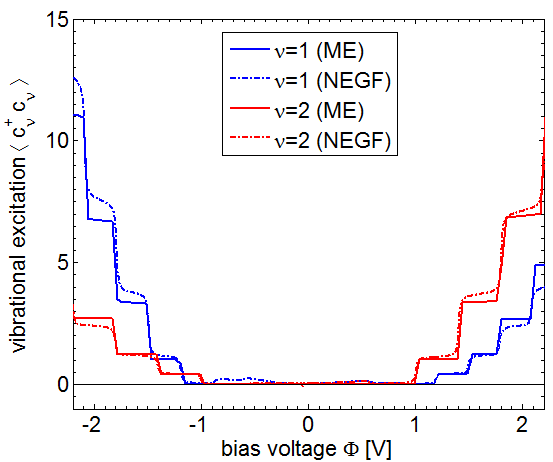}
}
\end{center}
  \caption{(Color online) Average levels of vibrational excitation for the two modes in model A as a function of the applied bias voltage $\Phi$. The blue and the red line depict the excitation characteristics of mode 1 and mode 2, respectively. Despite the different frequencies of the modes, mode 1 is much higher excited than mode 2 for negative bias voltages, while it is less excited than mode 2 for positive bias voltages. The excitation of the two modes can thus be selectively controlled by the external bias voltage $\Phi$ (MSVE).}\label{Fig.9c}
\end{figure}

\begin{figure}
\begin{center}
\resizebox{\newwidth}{\newheight}{
\includegraphics{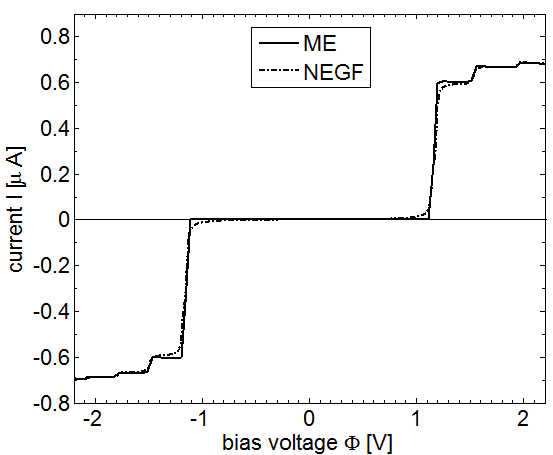}
}
\resizebox{\newwidth}{\newheight}{
\includegraphics{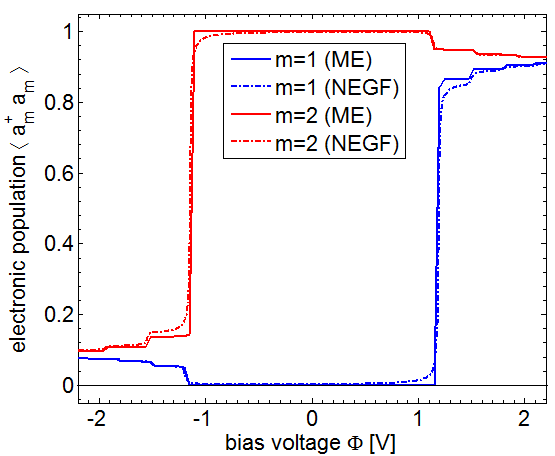}
}
\end{center}
  \caption{(Color online) Current-voltage characteristics and the respective population the electronic states for the model system depicted by Fig.\ \ref{Fig.8}b (model B). Solid (dashed dotted) lines are obtained employing the ME (NEGF) methodology. The population characteristics of state 1 and 2 are depicted by the blue and the red line,  respectively.  While the current is almost anti-symmetric with respect to the bias voltage $\Phi$, the population of the electronic levels reflects the asymmetry in the coupling of the molecule to the left and the right lead.}\label{Fig.10}
\end{figure}

\begin{figure}
\begin{center}
\resizebox{\newwidth}{\newheight}{
\includegraphics{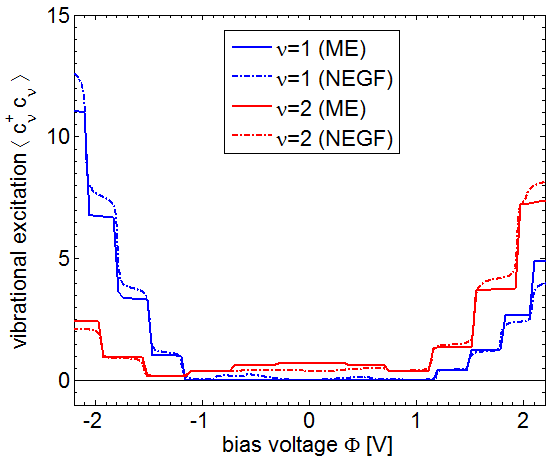}
}
\end{center}
  \caption{(Color online) Average levels of vibrational excitation for the two modes in model B as a function of the applied bias voltage $\Phi$. The blue and the red line depict the excitation characteristics of mode 1 and mode 2,  respectively. Despite the different frequencies of the two modes, mode 1 is much higher excited than mode 2 for negative bias voltages, while it is less excited than mode 2 for positive bias voltages. The excitation of the two modes can thus be selectively controlled by the external bias voltage $\Phi$ (MSVE).}\label{Fig.10c}
\end{figure}

First, we discuss results, where we do not account 
for a coupling between the vibrational modes 
and a thermal bath ($\zeta_1=\zeta_2=0$), nor 
for off-diagonal electronic-vibrational coupling ($\lambda_{\nu,m}\sim \delta_{\nu,m} $) 
or electron-electron interactions ($U_{1,2}=0$). 
Current-voltage characteristics and the corresponding population of the electronic states are shown in 
Figs.\ \ref{Fig.9} and \ref{Fig.10} for model A and B, respectively. The corresponding levels of 
vibrational excitation are depicted in Figs.\ \ref{Fig.9c} and \ref{Fig.10c}. 
In both models, the excitation of the two normal modes is calculated 
with respect to the neutral molecule, in which the two electronic states (LUMO and 
LUMO+1) are unoccupied. Notice, however, that while in model A these states become 
occupied only for non-zero bias, in model B the molecule (LUMO) is charged 
(and therefore to some extent vibrationally excited) 
already at zero bias.  
Since $\lambda_{\nu,m}\propto \delta_{\nu,m} $ and $U_{1,2}=0$, 
the two subsystems, the one comprising state 1 and mode 1,  
and the other consisting of state 2 and mode 2,  
are interrelated only by the coupling of the two electronic states 
to the leads. 
This coupling, however, does 
not induce strong correlations between the two subsystems, 
because the electronic states are non-degenerate, i.e., 
$\vert\epsilon_{2}-\epsilon_{1}\vert>\Gamma$. 
The transport characteristics of model A and B can therefore, in principle, be understood 
by the arguments given in Sec.\ \ref{VSE} for a single 
electronic state and a single vibrational mode. 
In particular, 
the agreement of the results obtained with NEGF and ME, which we already found in 
Sec.\ \ref{VSE}, is maintained.



While for both models the current-voltage characteristics is almost 
anti-symmetric with respect to the applied bias voltage $\Phi$, 
the electronic population and average levels of vibrational excitation 
exhibit strong asymmetric behavior. 
In particular, for negative bias voltages 
mode 1 shows a much higher level of vibrational excitation than mode 2. 
For positive bias voltages, however, the distribution of vibrational energy is reversed and 
mode 2 is higher excited than mode 1, despite the fact that $\Omega_{2}>\Omega_{1}$.

These results, where no intra-molecular interactions are considered, are in line with 
the interpretation and the analysis for cooling of vibrational modes 
by electron-hole pair creation processes (cf.\ Sec.\ \ref{VSE}). 
In particular, it is sufficient to consider the  
asymmetry in the coupling of each electronic state to the two leads, 
and the energy gap between each electronic state and the chemical potential 
of the two electrodes in order 
to assess which of the two vibrations is more effectively excited. 
In a realistic model of a molecular junction, however, 
correlations need to be taken into account. 
To this end, we analyze MSVE in the next three sections in terms of 
off-diagonal electronic-vibrational coupling, $\lambda_{\nu,m}\neq \delta_{\nu,m}$, 
electron-electron interactions, $U_{1,2}\neq0$, 
and in the presence of 
efficient cooling by coupling to a cold nuclear bath, 
 $\zeta_1\neq0$ and $\zeta_2\neq0$.

\subsubsection{MSVE in the Presence of Off-Diagonal Electronic-Vibrational Coupling}

\begin{figure} 
\begin{center}
\resizebox{\newwidth}{\newheight}{
\includegraphics{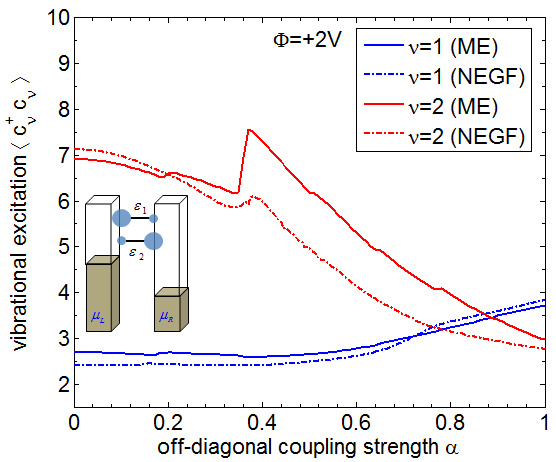}
}
\resizebox{\newwidth}{\newheight}{
\includegraphics{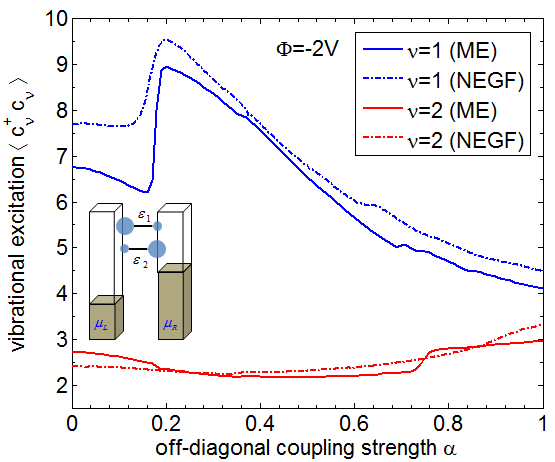}
}
\end{center}
  \caption{(Color online) Average vibrational excitation of the two vibrational modes in model A as a function 
of the off-diagonal electronic-vibrational coupling strength $\alpha$. 
The top and the bottom plots correspond to a fixed bias voltage of $\Phi=\pm2$\,V, respectively,  
as illustrated in the insets. Although off-diagonal electronic-vibrational coupling 
distributes current-induced excitation among the vibrational modes, 
MSVE occurs for a broad range of coupling strengths $\alpha$. }\label{Fig.11}
\end{figure}

\begin{figure} 
\begin{center}
\resizebox{\newwidth}{\newheight}{
\includegraphics{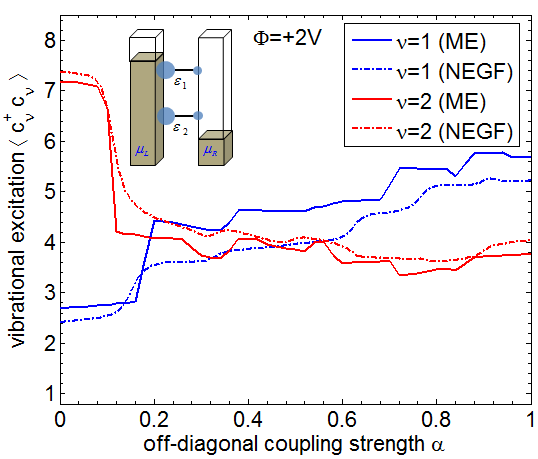}
}
\resizebox{\newwidth}{\newheight}{
\includegraphics{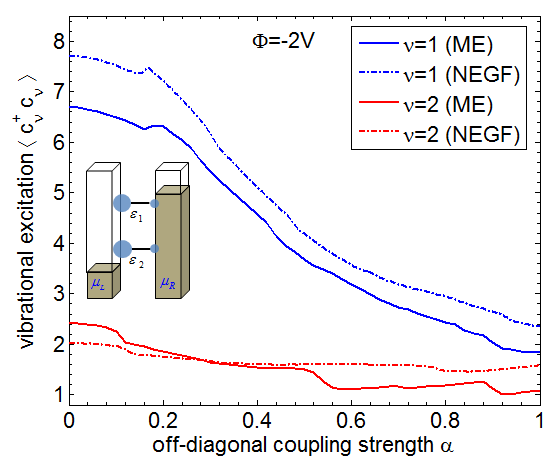}
}
\end{center}
  \caption{(Color online) Average vibrational excitation of the two vibrational modes in model B as a function 
of the off-diagonal electronic-vibrational coupling strength $\alpha$. 
The top and the bottom plots correspond to a fixed bias voltage of $\Phi=\pm2$\,V, respectively,  
as illustrated in the insets. Although off-diagonal electronic-vibrational coupling 
distributes current-induced excitation among the vibrational modes, 
MSVE occurs for a broad range of coupling strengths $\alpha$. }\label{Fig.12}
\end{figure}

As shown above, 
the MSVE phenomenon depends predominantly on 
the efficiency of cooling by electron-hole pair creation processes. 
This efficiency can be selectively controlled by the external bias voltage 
due to the asymmetry not only in the molecule-lead coupling, 
but also in the electronic-vibrational coupling.
The latter is most pronounced when each mode is coupled exclusively 
to a different electronic state, \emph{i.e.}, 
$\lambda_{\nu,m}\sim \delta_{\nu,m}$. 
In Figs.\ \ref{Fig.11} and \ref{Fig.12}, for model A and B, respectively, 
the vibrational excitation of the two modes 
is shown for increasing off-diagonal electronic-vibrational coupling: 
$\lambda_{1,2}=\alpha\lambda_{1,1}$ and $\lambda_{2,1}=\alpha\lambda_{2,2}$, 
where $\alpha=0$ describes the absence of off-diagonal vibronic coupling, while 
for $\alpha=1$ off-diagonal coupling is as strong as the diagonal one. 
Thereby, we use a fixed bias voltage $\Phi=+2$\,V 
for Figs.\ \ref{Fig.11}a and \ref{Fig.12}a, 
and $\Phi=-2$\,V for Figs.\ \ref{Fig.11}b and \ref{Fig.12}b. 
The results demonstrate that 
off-diagonal electronic-vibrational coupling 
tends to decrease MSVE for these model molecular junctions, 
as might have been anticipated. 
MSVE, however, remains significant 
for a broad range of coupling strengths $\alpha$. 
A more detailed analysis rationalizes the trends in each case. 

\begin{figure} 
\begin{center}
\includegraphics[scale=0.75]{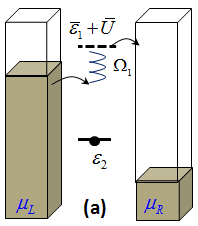}
\includegraphics[scale=0.75]{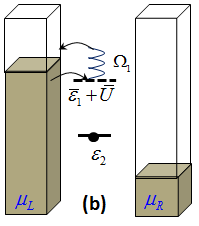}
\includegraphics[scale=0.75]{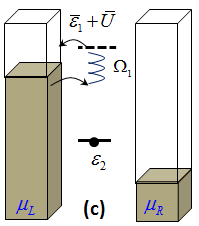}
\end{center}
 \caption{(Color online) Schematic representation of Coulomb-assisted transport processes (Panel (a)) and Coulomb-assisted electron-hole pair creation processes (Panels (b) and (c)). The dashed line represents the position of the di-anionic (doubly occupied) resonance, where state 2 is considered occupied. Similar processes are available at energies $\bar{\epsilon}_{2}+\bar{U}$. The position of the two electronic states with respect to the chemical potentials in the left and the right lead,
$\bar{\epsilon}_{1/2}-\mu_{\text{L/R}}$, and  the positions of the di-anionic resonances,  $\bar{\epsilon}_{1/2}+\bar{U}-\mu_{\text{L/R}}$, determine the efficiency of heating and cooling processes.}\label{Fig.17}
\end{figure}

In model A, at positive bias (Fig.\ \ref{Fig.11}a) and for $\alpha=0$, 
cooling by electron-hole 
pair creation is more effective via 
the state that is coupled more strongly to the left electrode (state 1). 
Therefore, the mode coupled to this state, that is mode 1, 
is more effectively cooled. 
As $\alpha$ increases, mode 2 becomes coupled to state 1, 
and cooling by electron-hole  
pair creation becomes effective also for this mode. 
While the level of excitation of mode 1 is thus almost the same for all values 
of $\alpha$, the one of mode 2 decreases. 
Similar arguments hold for negative bias voltages (Fig.\ \ref{Fig.11}b), 
where electron-hole pair 
creation via state 2 at the right electrode 
is the dominant cooling mechanism. \\
Notice that off-diagonal coupling also involves a change 
in the nuclear reorganization energy of 
each electronic state. This is particularly pronounced for transport 
and pair creation processes involving the di-anionic states 
(or a doubly occupied molecular bridge, 
cf.\ Fig.\ \ref{Fig.17}), 
which reorganization energies also involve  
vibrationally induced electron-electron interactions  
($\bar{U}_{1,2}\approx0.25$\,eV for $\alpha=1$, see Sec.\ \ref{NEGFapproach}). 
Since state 1 (2) is almost fully occupied for 
$\Phi=2$\,V ($\Phi=-2$\,V), 
processes involving state 2 (1) are dominated by 
the di-anionic resonance at 
$\bar{\epsilon}_{2(1)}+\bar{U}_{1,2}$. 
Increasing $\alpha$ shifts 
this resonance to significantly lower energies. 
This shift manifests itself in 
the kink observed in the vibrational excitation of mode 2 (1) 
at $\alpha=0.4$ for $\Phi=2$\,V (at $\alpha=0.2$ for $\Phi=-2$\,V),  
indicating the suppression of electron-hole pair creation processes 
(cf.\ Fig.\ \ref{Fig.17}a)  
as the respective resonance is shifted further away 
from the chemical potential in the left (right) 
electrode. 
Note that such kinks are less pronounced 
for larger values of $\alpha$, where cooling by electron-hole pair creation processes 
occurs for each mode 
via both electronic states such that 
the closure of one of these cooling channels 
is less significant.

Similar trends are observed for model B (cf.\ Fig.\ \ref{Fig.12}).
For negative bias and $\alpha=0$, mode 2 is more effectively cooled due to the strong 
coupling between state 2 and the left electrode. Increasing $\alpha$, mode 
1 becomes coupled to that state as well, leading to a suppression of vibrational excitation 
also for this mode. 
Notice that di-anionic resonances are less important in this case as the 
two states are almost unoccupied. Therefore, kinks associated with the 
reorganization energy of the electronic levels 
are also less pronounced. 
For positive bias, however, 
the two electronic states are almost fully occupied, 
and therefore, transport and pair creation processes do occur predominantly 
by the di-anionic resonances. 
Since these resonances are shifted to lower energies with increasing 
$\alpha$, state 1 is effectively located further away 
from the chemical potential in the 
left electrode and state 2 
closer to the one in the right electrode. 
This results in less (more) efficient 
cooling by electron-hole pair creation processes, and respectively, 
in an increased (decreased) level of vibrational excitation. 
The latter trends bring the excitation levels of the two modes to similar values 
already for $\alpha\approx 0.2-0.5$, which suppresses 
MSVE for this model. We note however, that the  
'Coulomb-like' attraction term ($\bar{U}$),  
which dominates the suppression of MSVE,  
is typically compensated by repulsive electron-electron interactions, 
which, however, are not accounted for in the present model.

\subsubsection{MSVE in the Presence of Electron-Electron Interactions }

\begin{figure} 
\begin{center}
\resizebox{\newwidth}{\newheight}{
\includegraphics{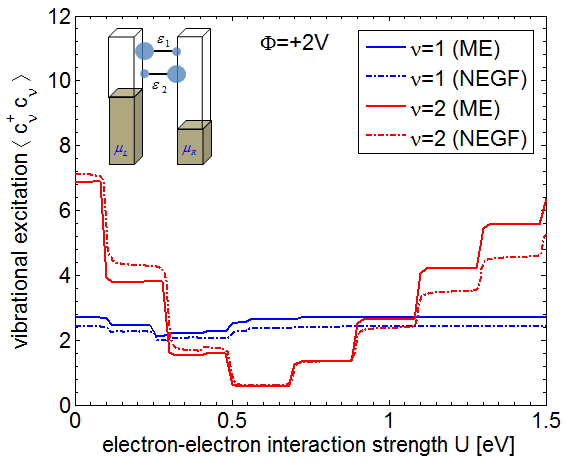}
}
\resizebox{\newwidth}{\newheight}{
\includegraphics{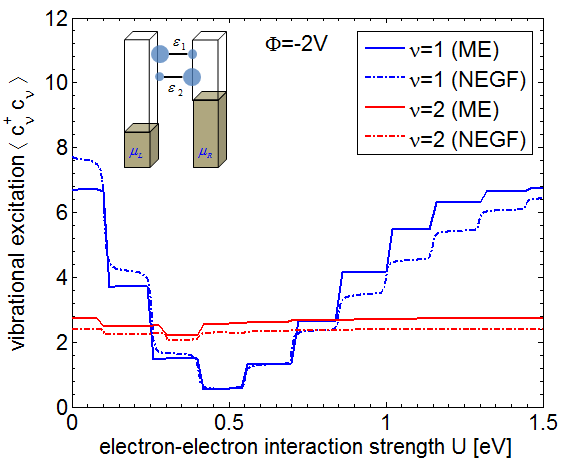}
}
\end{center}
  \caption{(Color online) Vibrational excitation as a function of the electronic interaction strength 
$U$ in model A. Blue (red) lines refer to the average excitation number of mode 1 (2). 
The top and the bottom plots correspond to a fixed bias voltage of $\Phi=\pm2$\,V, respectively. 
The efficiency of cooling mode 2 (1) by electron-hole pair creation processes with respect to the di-anionic (doubly occupied) state varies with the electron-electron interaction strengths $U$ 
for positive (negative) bias voltages. 
This leads to a inversion of MSVE for intermediate values of $U$, 
where the energy of the corresponding di-anionic state, $\bar{\epsilon}_{2}+U$ ($\bar{\epsilon}_{1}+U$), is close to the 
chemical potential in the left (right) lead.}\label{Fig.15}
\end{figure}

\begin{figure} 
\begin{center}
\resizebox{\newwidth}{\newheight}{
\includegraphics{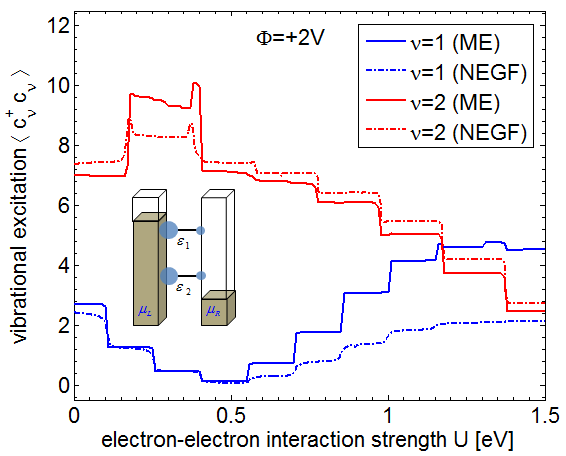}
}
\resizebox{\newwidth}{\newheight}{
\includegraphics{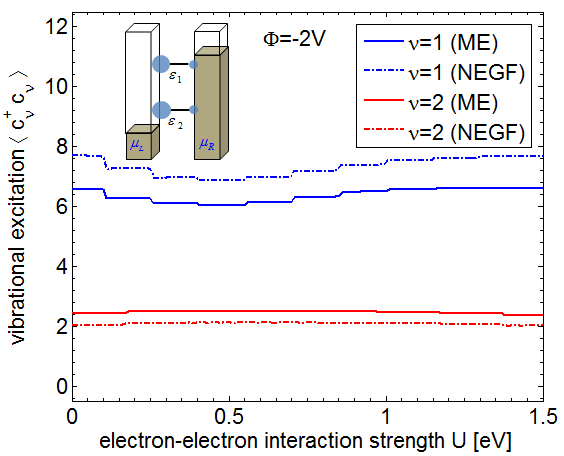}
}
\end{center}
  \caption{(Color online) Vibrational excitation as a function of the electronic interaction strength 
$U$ in model B. 
Blue (red) lines refer to the average excitation number of mode 1 (2). 
The top and the bottom plots correspond to a fixed bias voltage of $\Phi=\pm2$\,V, respectively. 
For negative bias voltages, due to the weak coupling to the right lead, 
both states are more or less unpopulated, 
which translates to a weak dependence of vibrational excitation on $U$. For positive bias voltages, however, 
double occupancy of the molecular bridge becomes important, and thus, the specific levels of vibrational excitation 
show a strong dependence on $U$. While MSVE thus exhibits a strong enhancement 
for weak and intermediate electron-electron interactions strengths, 
it is attenuated for higher values of $U$.}\label{Fig.16}
\end{figure}

A comparison of the electronic populations, shown in Figs.\ \ref{Fig.9}b and \ref{Fig.10}b, 
and the associated levels of vibrational excitation, given in Figs.\ \ref{Fig.9c} and \ref{Fig.10c}, 
shows that these quantities are strongly correlated. 
Electron-electron interactions, $U_{m,n}\neq0$, can strongly influence 
the electronic population of the different states 
\cite{Hettler2002,Hettler2003,Datta2007,Hartle2010b,Leijnse2011}
and thus the degree of MSVE 
in an asymmetric molecular junction. 
We study the effect of such inter-state correlations 
on MSVE in Figs.\ \ref{Fig.15} and \ref{Fig.16}, where the levels of excitation 
for the two vibrational modes in model A and B are plotted as functions of the 
electron-electron interaction strength $U_{1,2}=U$, 
using fixed bias voltages $\Phi=\pm2$\,V.


In model A (Fig.\ \ref{Fig.15}), the dependence of the two vibrational mode-excitations on $U$ is 
nearly reversed when the bias is reversed, demonstrating MSVE for most values of $U$. 
The particular levels of vibrational excitation and the direction of MSVE reflect the detailed 
asymmetry of the molecular junction. 
For example, for $\Phi=2$\,V the excitation of mode 1 is nearly 
independent on $U$, while the one of mode 2 shows a strong non-monotonic dependence. 
Since mode 1 is coupled to state 1, and because 
state 2 is almost unoccupied in this regime, 
mode 1 is efficiently cooled by electron-hole pair creation processes 
with respect to the left electrode, regardless of the electron-electron interaction strength $U$. 
In contrast, mode 2 is coupled to state 2 that is only weakly coupled to the left lead, 
and therefore, not efficiently cooled by electron-hole pair creation processes for $U=0$. 
However, as $U$ increases, electron 
tunneling with respect to state 2 involves an increasingly larger charging energy,  $\bar{\epsilon}_{2}+U$, 
since state 1 is almost fully occupied. This brings the 
electronic energy in this transport channel first closer to and then further away from the 
chemical potential in the left electrode. Cooling 
of mode 2 by electron-hole pair creation processes 
(Coulomb-assisted electron-hole pair creation processes as depicted in Figs.\ \ref{Fig.17}b and \ref{Fig.17}c)
is thus first enhanced and then suppressed as $U$ increases,  
resulting in the observed non-monotonic level of vibrational excitation. 
Notice that 
the pronounced cooling due to Coulomb-assisted electron-hole pair creation processes,   
in the intermediate regime of electron-electron interaction strengths, $U$, 
reverses the direction of MSVE with respect to the $U=0$ case 
(cf.\ Fig.\ \ref{Fig.15}).

In model B (Fig.\ \ref{Fig.16}) the asymmetry in 
the coupling of the molecular bridge to the electrodes 
leads to very different dependencies 
of the vibrational mode-excitations on $U$ for different polarities of the applied bias voltage. 
The resulting dependence of MSVE on $U$ is non-trivial, 
ranging from enhancement to suppression of the effect with respect to $U=0$.  
At negative bias voltages the two electronic states remain nearly unoccupied, 
so that the effect of electron-electron interactions is negligible. 
The lower excitation level of mode 2 at this polarity of the bias voltage 
reflects more efficient cooling by electron-hole pair creation via state 2 
at the left electrode (Fig.\ \ref{Fig.16}), 
which is maintained for different values of $U$. 
For positive bias voltages, the two electronic states are almost fully occupied. 
Cooling of mode 1 and 2 is thus dominated by electron-hole 
pair creation at the left and the right leads, respectively, 
according to the proximity of the corresponding energy levels, 
$\bar{\epsilon}_{1}+U$ and $\bar{\epsilon}_{2}+U$, 
to the respective chemical potentials. 
As $U$ increases, electron-hole pair creation at the left electrode is enhanced, while 
pair creation with respect to the right lead is suppressed. This 
leads first to an enhancement of MSVE with respect to $U=0$.  
At $U\approx0.4$\,eV, however,   
the electronic energy $\bar{\epsilon}_{1}+U$ crosses 
the chemical potential in the left lead, and state 1 
becomes discharged. 
At this point, cooling of mode 1 is at its maximal efficiency.
Simultaneously, since state 1 is no longer populated 
($\langle a_{1}^{\dagger} a_{1} \rangle$ drops from $\approx0.9$ to $\approx0.1$), 
processes involving state 2 
occur predominantly via the anionic channel at $\bar{\epsilon}_{2}$, 
such that cooling of mode 2 by electron-hole pair creation processes with 
respect to the right electrode becomes as efficient as for $U=0$. 
Accordingly, the level of excitation for mode 2 
returns to its original value. 
Increasing $U$ even further 
cooling of mode 1 (via the di-anionic channel, $\bar{\epsilon}_{1}+U$) 
becomes less efficient, 
and the level of excitation for mode 1 increases again.  
As $\bar{\epsilon}_{2}+U$ is closer 
to the chemical potential in the left lead, 
the cooling efficiency of mode 2 by electron-hole pair creation processes with 
respect to the left lead increases.    
The overall effect leads to a suppression of MSVE for $U\gtrsim1$\,eV in the given range of 
electron-electron interactions strengths $U$. 
Note, however, that for yet larger values of the electron-electron interaction strength, 
$U\gtrsim2.5$\,eV, MSVE is regained 
and becomes approximately as pronounced as for $U=0$.
In this regime, transport and pair creation processes
are dominated by the anionic resonances (at $\bar{\epsilon}_{1/2}$) so that 
the asymmetry in the cooling efficiency of the two modes (at least for the present model) 
is the same as for $U=0$.

\subsubsection{MSVE for Strong Vibrational Relaxation}

In the discussion of Fig.\ \ref{Fig.7} in Sec.\ \ref{VSE}, we have already seen that the level of excitation 
of a single vibrational mode consists of two contributions: current-induced local heating due 
to inelastic electron transport processes (as shown in Figs.\ \ref{basmech}a-d), 
and polaron-formation \cite{Thoss2011}, which can be quantified by the difference 
\begin{eqnarray}
 \langle c_{\nu}^{\dagger}c_{\nu} \rangle_{\hat{H}}-\langle c_{\nu}^{\dagger}c_{\nu} \rangle_{\bar{H}}=\sum_{mm'}(\lambda_{\nu,m}\lambda_{\nu,m'}/\Omega_{\nu}^{2}) \langle a_{m}^{\dagger}a_{m} a_{m'}^{\dagger}a_{m'} \rangle_{\bar{H}}. 
\end{eqnarray}
Strong vibrational relaxation results in a strong suppression of 
current-induced vibrational excitation, especially if 
the time-scale for vibrational relaxation 
is much shorter than the time-scale between two consecutive 
transport events. 
The contribution due to polaron formation, however, stems 
from the steady-state population of the electronic 
levels in a molecular junction.  
Hence, for model A, 
MSVE occurs in the presence of strong vibrational relaxation \cite{MSVE10}, 
since the population of the electronic levels can 
be selectively controlled by the external bias voltage $\Phi$ (cf.\ Fig.\ \ref{Fig.9}b). 
In model B, however, both states are either fully populated or 
empty such that strong vibrational relaxation 
is likely to hinder MSVE for this model system.

\section{Conclusion}

In this work we have studied and analyzed transport characteristics of single-molecule junctions, focusing on the excitation of specific molecular vibrational modes. In particular, we have shown that the level of excitation of specific modes can be controlled by the polarity and the magnitude of an external bias voltage. Thereby, high-frequency modes (typically associated with strong chemical bonds) can be higher excited than low-frequency modes, which translates to a "non-statistical" distribution of energy among the vibrational modes. We refer to this phenomenon as mode-selective vibrational excitation. 

Our main findings are summarized below:
\begin{itemize}
 \item[1)] \emph{The importance of cooling by electron-hole pair creation}\\
Our analysis shows that cooling of the vibrational modes in a molecular junction 
by electron-hole pair creation processes is crucial to understand the extent of the MSVE phenomenon. 
In particular, since the efficiency of these processes is sensitive to the position 
of the chemical potentials in the leads, the levels of vibrational excitation in the molecule 
can be controlled by an external bias voltage. Considering a molecule with multiple vibrational modes 
and typical asymmetries in the vibronic as well as molecule-lead couplings,  
the level of excitation of specific vibrational modes can thus be tuned by 
the polarity and the magnitude of the external bias voltage.  
 \item[2)] \emph{The role of asymmetry and intra-molecular interactions}\\ 
Our studies suggest that MSVE is a rather general phenomenon and is likely to be observed experimentally. 
The required asymmetry in the electronic interaction between different molecular states and the 
leads may be due to an inherent asymmetric 
molecular structure (model A) or stem from an inherent difference between the two electrodes, 
as \emph{e.g.}\ in STM experiments (model B). 
Intra-molecular interactions, \emph{e.g.}\ due to off diagonal electronic-vibrational coupling or 
electron-electron repulsion tend to redistribute the excitation energy between 
the different modes, and thus work against MSVE. However, 
having analyzed a broad range of parameters, we found the MSVE phenomenon 
to prevail even in the presence of 
such interactions. 
 \item[3)] \emph{The importance of off-resonant processes: Comparing ME to NEGF calculations.}\\ 
Our numerical studies of generic models of molecular junctions were based on two 
complementary theoretical methods: a nonequilibrium Green's function 
approach \cite{Hartle,Hartle09,MSVE10,Hartle2011b} and a master equation 
approach \cite{Volkovich2008,Peskin2010,Volkovich2011}. Both approaches are based 
on a second-order expansion in the coupling of the molecular bridge to the leads. 
While the NEGF method also accounts for higher-order effects, the ME approach 
describes only resonant electron tunneling processes. However, intra-molecular 
interactions, either due to electronic-vibrational coupling or electron-electron interactions, 
can be described by the ME method without invoking further approximations, while 
these interactions are described by our NEGF approach approximately 
in terms of a non-perturbative scheme. Although the results obtained 
by both methodologies agree in most cases reasonably well, further insights 
into the relevant mechanisms can be gained when the results exhibit differences. 
Thus, for example, the importance of off-resonant electron-hole pair creation 
processes for local cooling \cite{Hartle09,Saito2009,Schiff2010,Romano10,Hartle2010b} 
of vibrational modes in the high-bias regime could be revealed. 
\end{itemize}

We end by noting that this work considered only  generic models to study the basic 
mechanisms and prerequisites of bias-controlled MSVE. The identification of 
specific molecules that exhibit MSVE requires transport studies based on 
first-principles electronic structure calculations 
\cite{Pecchia04,Frederiksen04,Benesch06,Troisi2006,Benesch06,Benesch08,Benesch09,Monturet2010}. 
This will be the subject of future work. 
Another interesting extension concerns the external control mechanism for MSVE. In the present work, 
we have considered an external bias voltage as the means to control MSVE. 
A gate electrode \cite{Sapmaz05,Song2009,Osorio2010,Martin2010} may provide another tool 
for addressing a molecular junction with an electric field, and thus, may also be used 
to control vibrational excitation. Finally, in the context of mode-selective chemistry, 
studies of MSVE in single-molecule junctions may pave the way to control chemical 
processes in molecules adsorbed on surfaces. For example, an additional electrode 
that provides an external potential bias may induce catalytic 
reactions in a selective manner.

\emph{Acknowledgements:}

We gratefully acknowledge fruitful discussions with O.\ Godsi, D.\ Brisker Klaiman, 
M.\ Butzin, P.\ Brana-Coto and O.\ Rubio-Pons. 
This research was supported by the German-Israeli Foundation for 
scientific development (GIF). RV acknowledges support from the Gutwirth Foundation. 
The Leibniz Rechenzentrum Munich (LRZ) and the Regionales Rechenzentrum Erlangen (RRZE) 
were providing the computer resources for our studies. 
The work at the Friedrich-Alexander Universit\"at Erlangen-N\"urnberg was carried out in the 
framework of the Cluster of Excellence "Engineering of Advanced Materials".

\end{document}